\documentclass[a4paper]{JHEP3} 
\usepackage{subfig}
\usepackage{microtype}
\usepackage{lmodern}
\usepackage[T1]{fontenc}
\usepackage{amssymb}
\usepackage{mathptmx}
\usepackage{graphicx}
\usepackage{color}
\let\normalcolor\relax 
\definecolor{myred}{rgb}{0.6,0,0} 
\definecolor{myblue}{rgb}{0,0.2,0.4}
\definecolor{mygreen}{rgb}{0,0.9,0.1}
\definecolor{hc}{rgb}{.9,0.1,0.7}
\definecolor{hcout}{rgb}{.9,0.7,0.9}
\definecolor{Orange}{rgb}{1.,0.65,0.}
\usepackage{amsmath}
\usepackage{rotating}
\usepackage{textcomp}
\usepackage{timestamp} 

\numberwithin{equation}{section}
\numberwithin{figure}{section}
\numberwithin{table}{section}

\newcommand{\ssL}{{\scriptscriptstyle{L}}}
\newcommand{\ssN}{{\scriptscriptstyle{N}}}
\newcommand{\ssO}{{\scriptscriptstyle{O}}}
\newcommand{\ssQ}{{\scriptscriptstyle{Q}}}
\newcommand{\ssE}{{\scriptscriptstyle{E}}}
\newcommand{\ssD}{{\scriptscriptstyle{D}}}
\newcommand{\ssB}{{\scriptscriptstyle{B}}}
\newcommand{\ssY}{{\scriptscriptstyle{Y}}}

\newcommand{\be}{\begin{equation}}
\newcommand{\ee}{\end{equation}}
\newcommand{\bea}{\begin{eqnarray}}
\newcommand{\eea}{\end{eqnarray}}

\newcommand{\nl}{\nonumber \\}

\def \litwo {{\rm{Li_2}}}

\title{%
NNLO leptonic and hadronic corrections to Bhabha scattering
          and luminosity monitoring at meson factories
}

\author{%
C.~Carloni Calame~${}^{a}$,
H.~Czy\.z~${}^{b}$,
J.~Gluza~${}^{b}$,
M.~Gunia~${}^{b}$,
G.~Montagna~${}^{c,d}$,
O.~Nicrosini~${}^{d}$,
F.~Piccinini~${}^{d}$,
T.~Riemann~${}^{e}$,
M.~Worek~${}^{f}$

  \\
  $^{a}$~School of Physics and Astronomy, University of Southampton, Southampton SO17 1BJ, U.K.
  \\
  $^{b}$~
   Department of Field Theory and Particle Physics,
      Institute of Physics, 
      University of Silesia, Uniwersytecka 4, PL-40-007 Katowice,
      Poland
  \\
  $^{c}$~Dipartimento di Fisica Nucleare e Teorica, Universit\`a di Pavia, Via A. Bassi 6, I-27100 Pavia, Italy
  \\
  $^{d}$~INFN, Sezione di Pavia, Via A. Bassi 6, I-27100 Pavia, Italy
  \\
  $^{e}$~Deutsches Elektronen-Synchrotron, DESY, Platanenallee
    6, 15738 Zeuthen, Germany
  \\
  $^{f}$~Fachbereich C Physik, Bergische Universit\"{a}t Wuppertal, 
   Gaussstr. 20,
  D-42097 Wuppertal, Germany

\\

\email{carlo.carloni-calame@soton.ac.uk}
, \email{czyz@us.edu.pl},  \email{janusz.gluza@us.edu.pl}, \email{guniamichal@gmail.com}, \email{guido.montagna@pv.infn.it}, \email{oreste.nicrosini@pv.infn.it}, \email{fulvio.piccinini@pv.infn.it}, \email{tord.riemann@desy.de}, \email{worek@physik.uni-wuppertal.de}
}

\abstract{
 Virtual fermionic $N_f=1$ and $N_f=2$ contributions to Bhabha scattering are
 combined with realistic  real corrections at next-to-next-to-leading order in QED.
The virtual corrections are determined  by the package \textsc{bha\_nnlo\_hf}, and real corrections with the Monte Carlo generators \textsc{Bhagen--1Ph}, 
\textsc{Helac--Phegas} and \textsc{Ekhara}. 
Numerical results are discussed at  the energies of and with realistic cuts used at the $\Phi$ factory DA$\Phi$NE, 
at the $B$ factories PEP-II  and KEK, 
and at the charm/$\tau$ factory  BEPC II. 
We compare these complete calculations with the approximate ones realized in
the generator \textsc{BabaYaga@NLO}
used at meson factories to evaluate
their luminosities. For realistic reference event selections we find agreement for the NNLO
leptonic and hadronic corrections within 0.07\% or better and conclude that they are well
accounted for in the generator by comparison with the present experimental accuracy.
\color{black}
}

\keywords{Electromagnetic Processes and Properties, NLO Computations}

\preprint{DESY 11--080
\\
FNT/T 2011/01
\\
LPN--11--25
\\
SFB--CPP--11--26
\\
WUB/11--05
}

\begin{document}


\section{Introduction\label{sec-intro}}
The Bhabha scattering process
\begin{eqnarray}\label{eq-lo}
e^+e^- \to e^+e^-
\end{eqnarray}
is an invaluable tool for the luminosity
determination in various experiments.
Both low energy devices, operating from about 1 GeV to several GeV and high energy devices, planned to operate
at hundreds or thousands of GeV, require theoretical predictions for the Bhabha cross section with
quite accurate determinations of QED radiative corrections. The latter contain, besides exponentiated
leading logarithmic terms, also the complete fixed order contributions, and in particular the complete
two-loop QED corrections.

Aiming at per mille accuracy or slightly better, the radiative  corrections may neglect the constant terms in the electron mass $m_e$.
At next-to-leading order (NLO) final states with unresolved photons will contribute,
\begin{eqnarray}\label{eq-nlo}
 e^+e^- \to  e^+e^-(\gamma) .
\end{eqnarray}
Further, new  mass scales start to play a role, but only in the one-loop self-energy insertions.
At next-to-next-to-leading order (NNLO), further final states of Bhabha scattering contain 
unresolved photons, fermion pairs,  or  hadrons:
\begin{eqnarray}\label{eq-nnlog}
 e^+e^- &\to& e^+e^-(\gamma,\gamma\gamma),~~~  e^+e^-(e^+e^-),  ~~~ e^+e^-(f^+f^-),  ~~~ e^+e^-(\mathit{hadrons}) .
\end{eqnarray}
At this order of perturbation theory,
 a variety of Feynman diagrams depend on additional mass parameters, and
 one may formally distinguish between $N_f=1$ corrections (with only electrons) and $N_f=2$ corrections, and the latter ones are technically more complicated due to the additional mass scale.

The NLO corrections, with inclusion of certain leading higher order terms, are known since a while and several Monte Carlo (MC) programs are carefully tuned.
A recent comprehensive review on precision predictions for scattering experiments at meson factories contains a detailed discussion of the state of the art \cite{Actis:2010gg}.

As far as virtual corrections are concerned, in the last few years
there has been major progress in the evaluation of the
corrections at the NNLO accuracy.
In fact, the photonic
two-loop QED corrections were first evaluated in the massless case in
\cite{Bern:2000ie}.
The  photonic corrections to massive Bhabha scattering
with enhancing powers of  $\ln(s/m_e^2)$ were soon derived from that
\cite{Glover:2001ev}.
The missing constant term in $m_e$ \cite{penin:2005kf} plus the
corrections with electron loop insertions \cite{Bonciani:2003cj0,Bonciani:2004qt,Bonciani:2004gi,Actis:2007gi}, called the $N_f=1$ case,
followed not much later.
The heavy fermion (or $N_f=2$) corrections
were first derived in the limit $m_e^2<<m_f^2<<s,|t|,|u|$
\cite{Actis:2007gi,Becher:2007cu}, where $m_f$ is the mass of the heavy fermion and $s,t, u$ are the usual Mandelstam variables,
and soon after also for $m_e^2<<m_f^2,s,|t|,|u|$
\cite{Bonciani:2007eh,Actis:2007pn}.
Finally, using dispersion relations, also hadronic corrections became
known \cite{Actis:2007fs,Actis:2008br,Kuhn:2008zs}.

 A complete collection of all the relevant formulae for the massive
 virtual NNLO corrections used in this paper can be found in the just
 mentioned papers and in \cite{ACGR-bha-nnlo-ho:2011}.

When fermion loops or virtual hadronic corrections are taken into account, the question of considering also the real emission of the corresponding particles arises.
That was studied for the emission of electron pairs in \cite{Arbuzov:1995vj} in the soft limit of electron pair energy and  in logarithmic accuracy. 
It is shown that the leading logarithmic corrections $\ln^3(s/m_e^2)$ cancel with those from the irreducible two-loop vertex corrections with electron loops.
A similar cancellation is expected for the combination of heavy fermion pair emission with  irreducible two-loop vertex corrections with a heavy fermion loop.
In practice, however, the situation is evidently a bit more involved, especially at smaller energies, when $s, |t|, |u| \sim m_f^2$.
Then, the logarithms are not numerically dominating and more diagrams get important.
Nowadays, 
MC programs can do that job.
In this article, we will perform such a study of reaction (\ref{eq-nnlog}) due to the additional emission of real pairs of leptons
with the Fortran packages \textsc{Helac--Phegas} 
 \cite{Kanaki:2000ey,Papadopoulos:2000tt,Papadopoulos:2005ky,Cafarella:2007pc}, 
 similarly to what was done in the 1990s for small-angle Bhabha scattering at LEP \cite{Montagna:1998vb,Montagna:1999eu}. 
For heavy fermions and hadrons, we present here the corresponding results for the first time.
The case of real hadron emission in Bhabha scattering deserves special
attention.
The only existing event generator contains only  pion pair emission and
the results obtained for the real pion pair emission serve as an
indication of the size of other left over corrections.
For a consistent treatment, we replace the hadronic dispersion integrals in the virtual and soft real hadronic corrections as described in \cite{Actis:2007fs,Actis:2008br} by the pion pair form factor.
The prediction is then combined consistently with real pion pair
emission as evaluated with the
  MC event generator \textsc{Ekhara} \cite{Czyz:2010sp,Czyz:2006dm,Czyz:2005ab,Czyz:2003gb}.
In the calculations we use the pion form factor from \cite{Bruch:2004py}.
 For the virtual and hard photon corrections
 the full hadronic corrections were obtained using the vacuum
 polarisation insertions.

The results may be compared with those from the Bhabha generators which are usually applied for experimental simulations of Bhabha scattering; here we look at
BabaYaga~\cite{CarloniCalame:2000pz,CarloniCalame:2001ny,CarloniCalame:2003yt}, in particular to the latest and most accurate
version \textsc{BabaYaga@NLO}~\cite{Balossini:2006wc}.
\color{black}

The aim of the article is to put together all the above discussed NNLO
corrections to Bhabha scattering taking into account real
experimental conditions and examine how well they are accounted for
 in the event generator \textsc{BabaYaga@NLO} used at meson factories for their
 luminosity measurements.
So far, at NNLO level, virtual corrections have been checked in detail only
for situations where
 the dependence of soft radiation on the maximum soft photon energy $\omega$ (or, equivalently, the minimal hard photon energy)
  is ``switched off" by setting $\omega=\sqrt{s}/2$
 \cite{penin:2005kf,Actis:2007fs,Actis:2008br,Kuhn:2008zs}.
This was a good way to compare results obtained by different theoretical groups, but certainly has nothing to do with reality.
We restrict ourselves here to low energies (meson factories) because presently they are of immediate relevance from the experimental point of view.

We just mention for completeness the last remaining NNLO issue:
that of radiative loop corrections, i.e. the NNLO contributions from the interference of photonic bremsstrahlung off one-loop diagrams with lowest order real photon contributions, first studied in \cite{Melles:1996qa} with a restriction to the factorising diagrams.
The technical complications arise from non-factorising diagrams, the so-called pentagon diagrams.
Recent papers on this issue are  \cite{Fleischer:2007ph,Kajda:2009aa,Actis:2009zz,Actis:2009uq}, but so far without explicit numerical results.
Sample numbers for
the virtual one-loop (plus real soft) QED corrections to the hard-bremsstrahlung process $e^+ e^- \to  e^+e^-\gamma$
are given in \cite{Actis:2009uq}. For future measurements, it would be worthwhile 
to answer the question if (and when) these corrections
 have to be yet included 
 in the MC event generators employed for simulating Bhabha scattering
events at low-energy high-luminosity electron-positron colliders.

The paper is organised as follows.
In Section \ref{sec-nnlo} we discuss the exact NNLO massive corrections
to Bhabha scattering and present benchmark results for event selections
close to the experimental ones. In Section \ref{sec-babayaga} we describe
the approximate treatment of these corrections in the \textsc{BabaYaga@NLO} event
generator and derive benchmark results for the same event
selections as for the exact results.
In Section \ref{sec-num} we show detailed numerical studies of the
quality of the approximations used in \textsc{BabaYaga@NLO}. We draw our
conclusions in Section \ref{sec-concl}.

\section{The NNLO massive corrections\label{sec-nnlo}}
The complete  NNLO $N_f=1,2$ corrections to Bhabha scattering consist of three parts, each of them with contributions from  virtual and real electron pair corrections ($N_f=1$ case) and corrections due to muon pairs, $\tau$ pairs and hadrons ($N_f=2$ cases):
\begin{eqnarray}\label{all-fermions}
\frac{d \sigma^{\rm{\ssN\ssN\ssL\ssO}}_{N_f}} {d\Omega}
&=&
\frac{d \sigma^{\rm{\ssN\ssN\ssL\ssO}}_{\mathit virt}} {d\Omega}
+
\frac{d \sigma^{\rm{\ssN\ssL\ssO}}_{\gamma}} {d\Omega}
+
\frac{d \sigma^{\rm{\ssL\ssO}}_{\mathit real}} {d\Omega}
\nonumber \\
&=&
\frac{d \sigma_{e^+e^-}} {d\Omega}
+
\frac{d \sigma_{\mu^+\mu^-}} {d\Omega}
+
\frac{d \sigma_{\tau^+\tau^-}} {d\Omega}
+
\frac{d \sigma_{\mathit{had}}} {d\Omega} .
\end{eqnarray}
We want to concentrate here on the interplay of virtual and real corrections.
For the various pure self-energy corrections in
$ \sigma^{\rm{\ssN\ssN\ssL\ssO}}_{\mathit virt}$
we refer to \cite{Actis:2010gg,Actis:2008br} and the references quoted therein.
This is in accordance with the approach chosen in the MC packages used for the interpretation of experimental results, and  we will not include these pure 
two-loop self-energy corrections  in the numerical results discussed below.

\begin{figure}[t]
\centering
\subfloat[][]{\includegraphics[width=.2\textwidth]{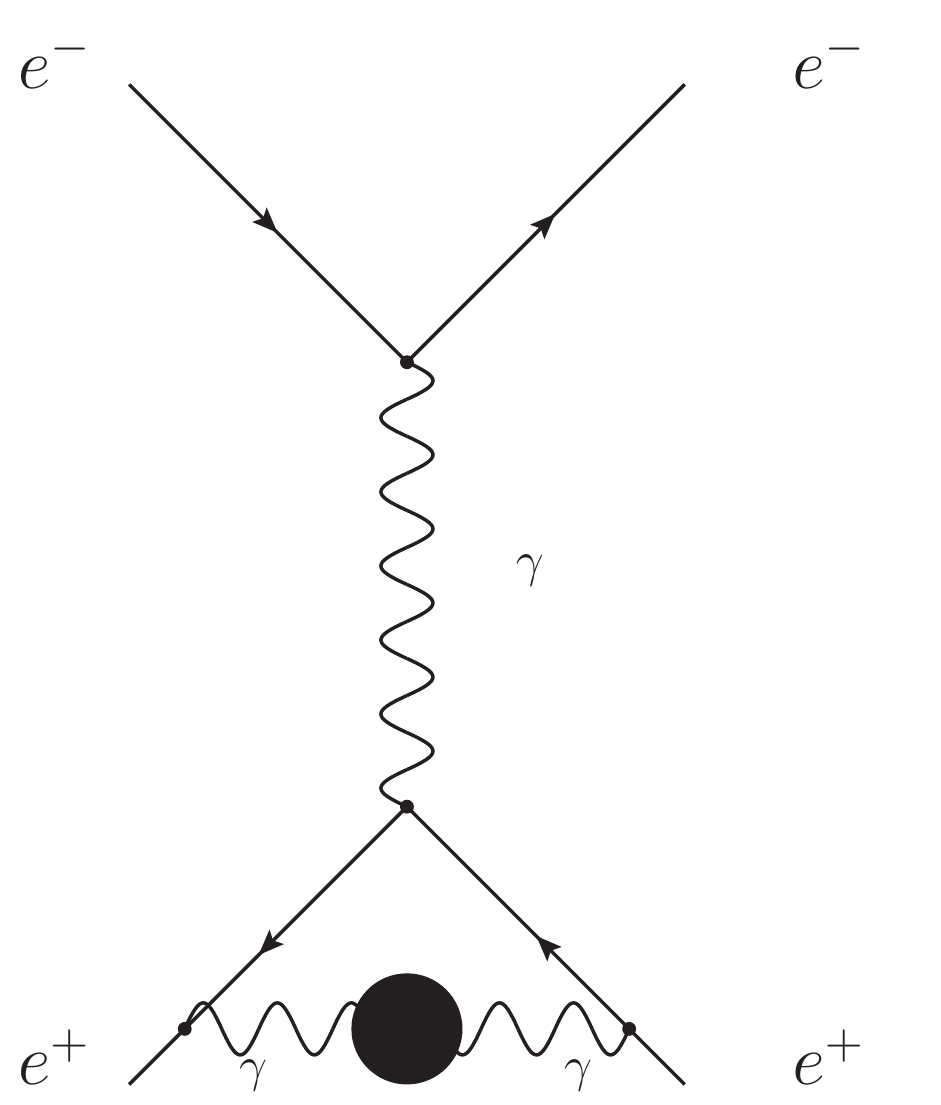}}
\hfill
\subfloat[][]{\includegraphics[width=.2\textwidth]{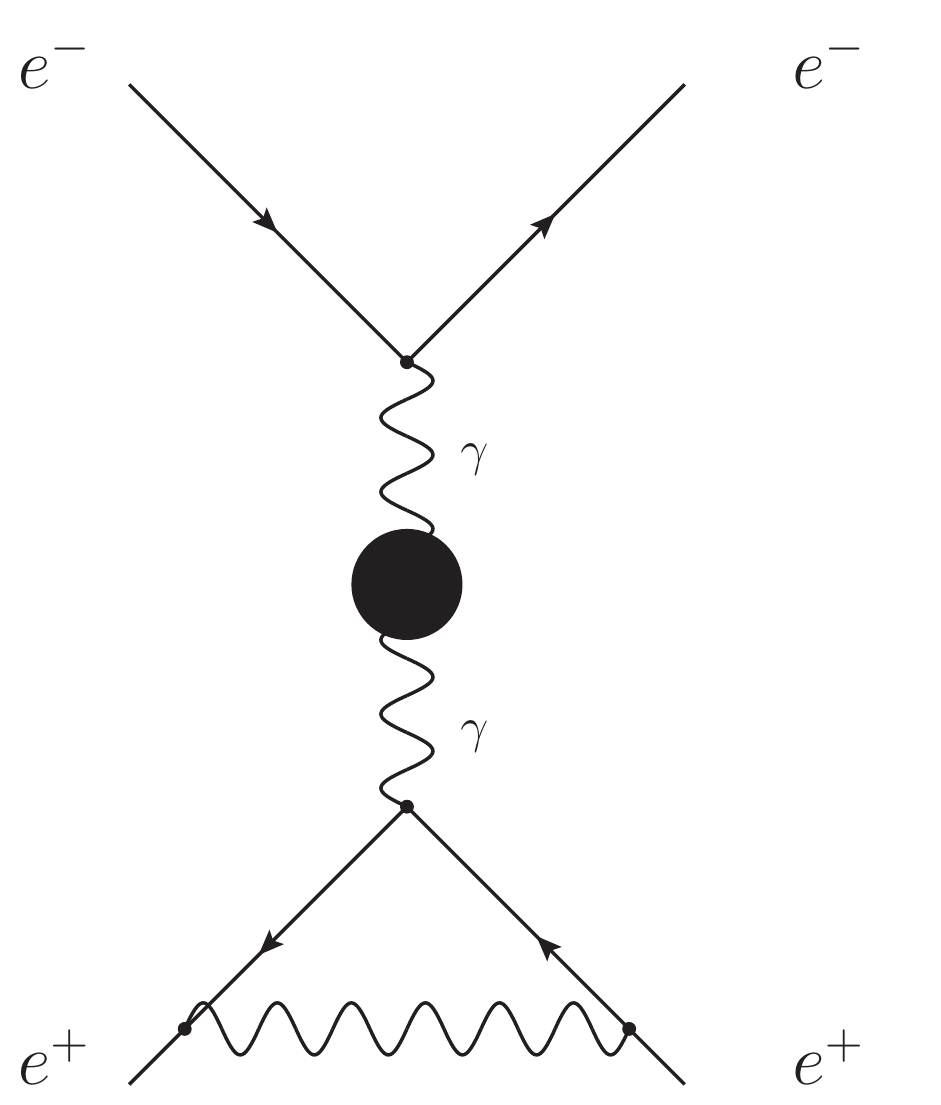}}
\hfill
\subfloat[][]{\includegraphics[width=.2\textwidth]{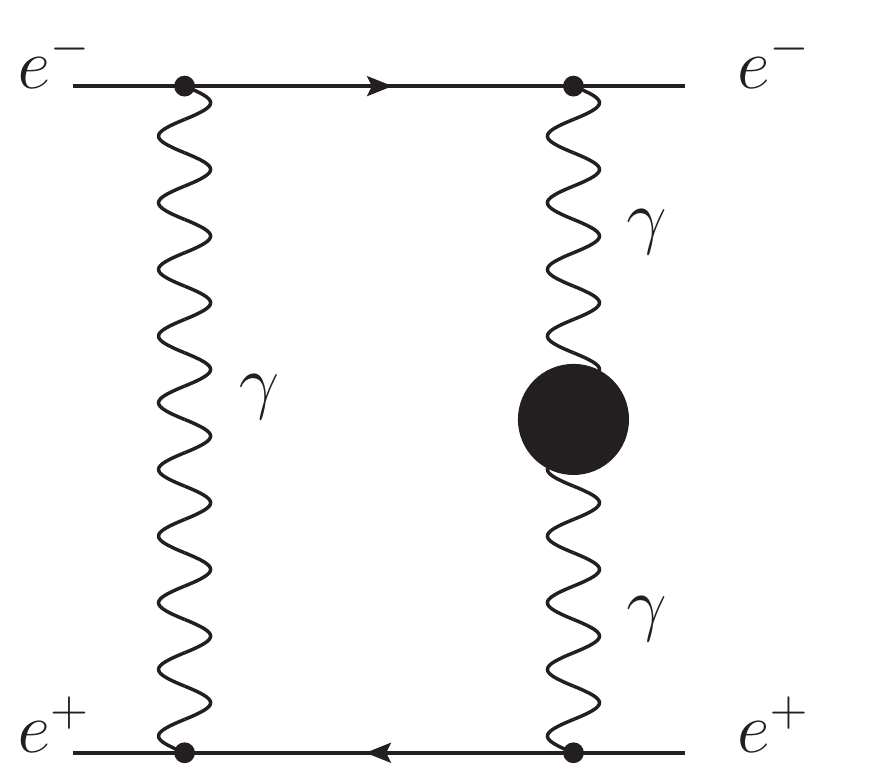}}
\hfill
\subfloat[][]{\rotatebox{90}{{\includegraphics[angle=180,width=.2\textwidth]{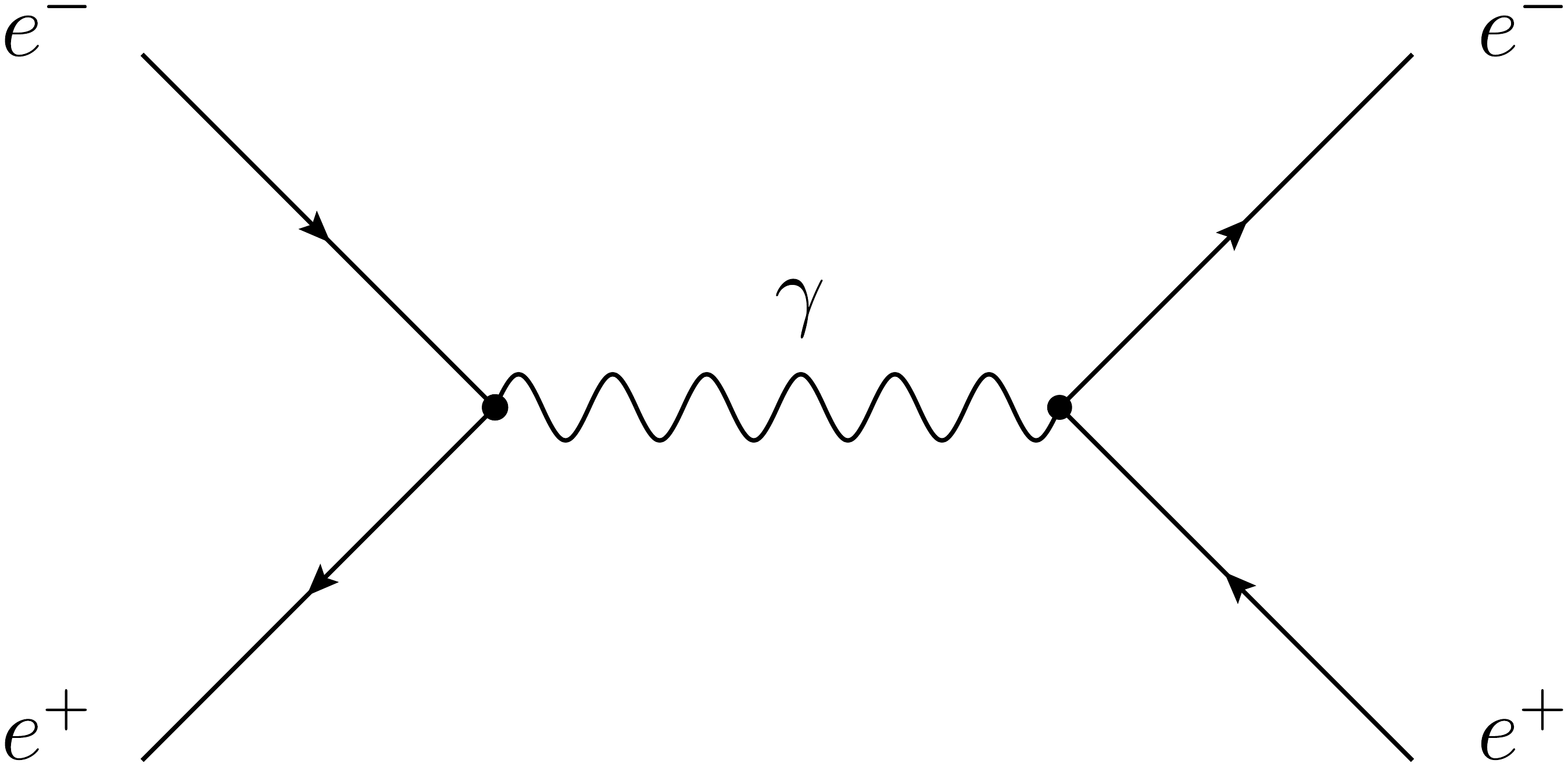}}}}
\caption[]{\emph{(a)--(c) are sample two-loop diagrams; their interference with (d) is contributing to
$\sigma^{\rm{\ssN\ssN\ssL\ssO}}_{\mathit 2L}$, part of
$\sigma^{\rm{\ssN\ssN\ssL\ssO}}_{\mathit virt}$, Eq.~(\ref{all-fermions}).}
}
\label{Ca}
\end{figure}

\begin{figure}[t]
\begin{center}
\subfloat[][]{\includegraphics[width=.2\textwidth]{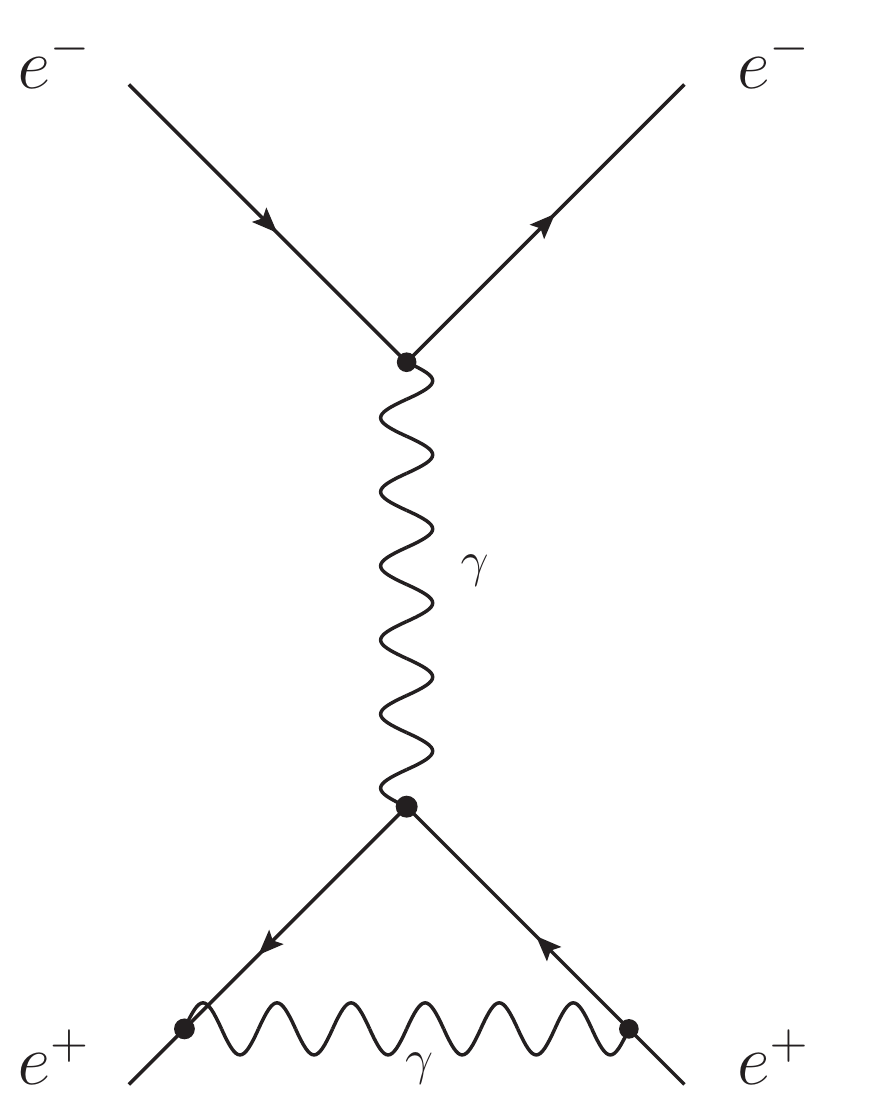}}
\subfloat[][]{\includegraphics[width=.2\textwidth]{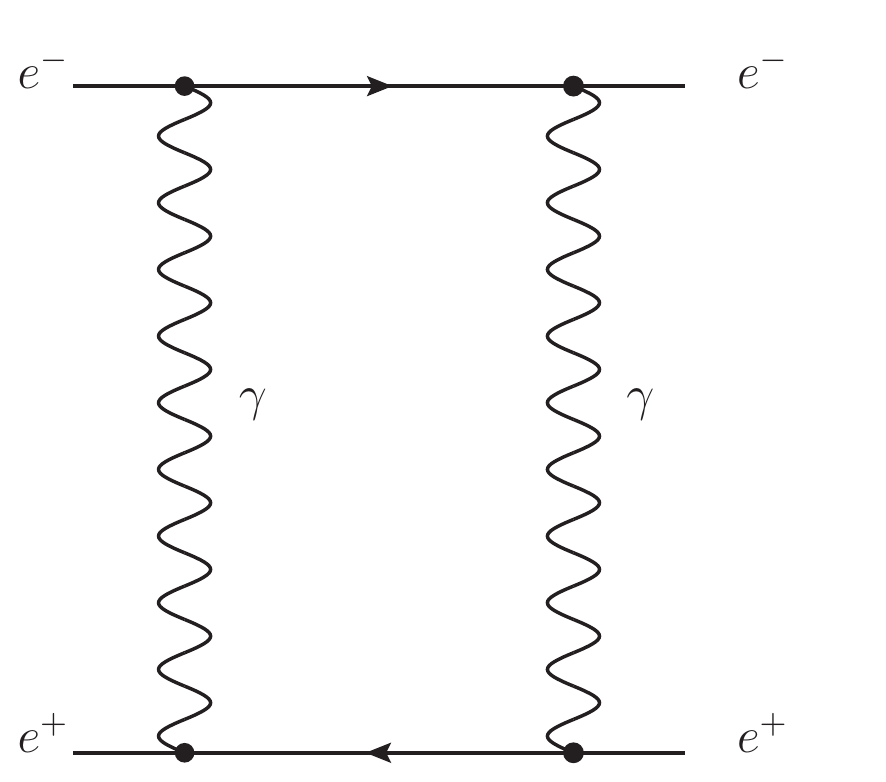}}
\subfloat[][]{\includegraphics[width=.2\textwidth]{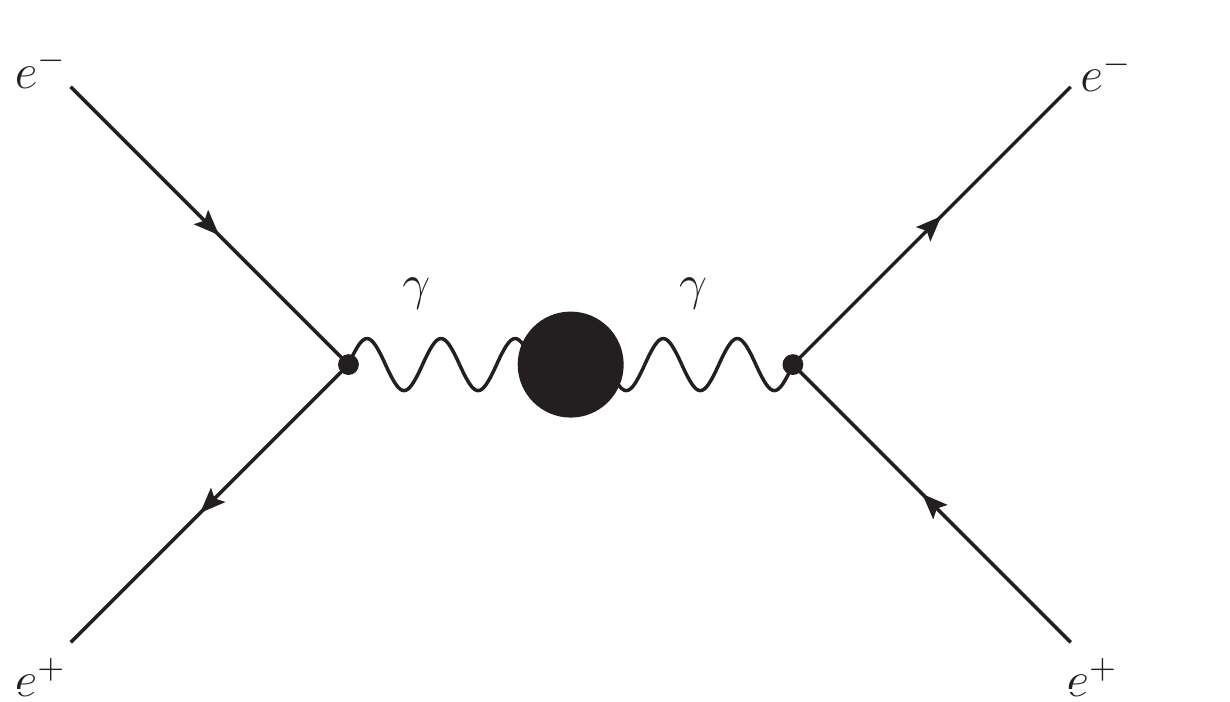}}
\end{center}
\caption[]{\emph{(a)--(b) are interfering with (c); they are samples of the so-called loop-by-loop corrections
$\sigma^{\rm{\ssN\ssN\ssL\ssO}}_{\mathit{1L1L}}$, part of
$\sigma^{\rm{\ssN\ssN\ssL\ssO}}_{\mathit virt}$, Eq.~(\ref{all-fermions}).}}
\label{Cb}
\end{figure}

\begin{figure}[b]
\begin{center}
\subfloat[][]{\includegraphics[width=.2\textwidth]{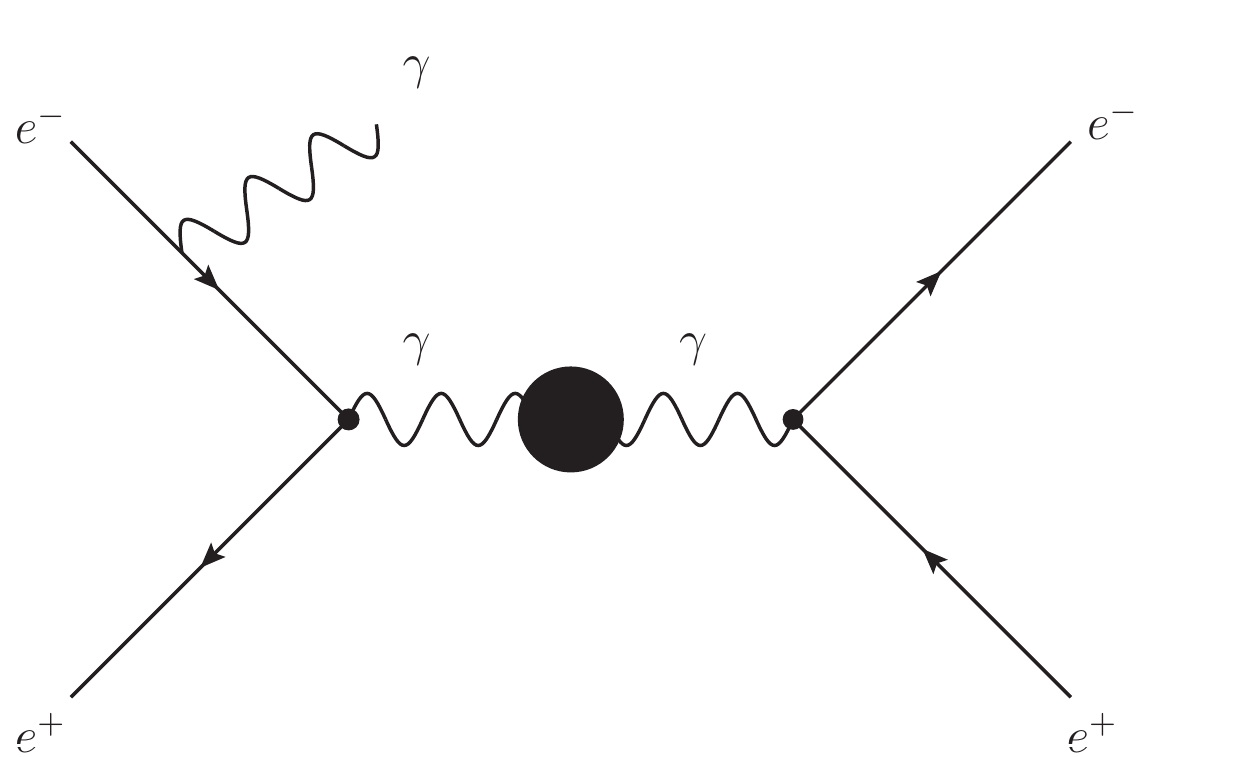}}
\subfloat[][]{\includegraphics[width=.25\textwidth]{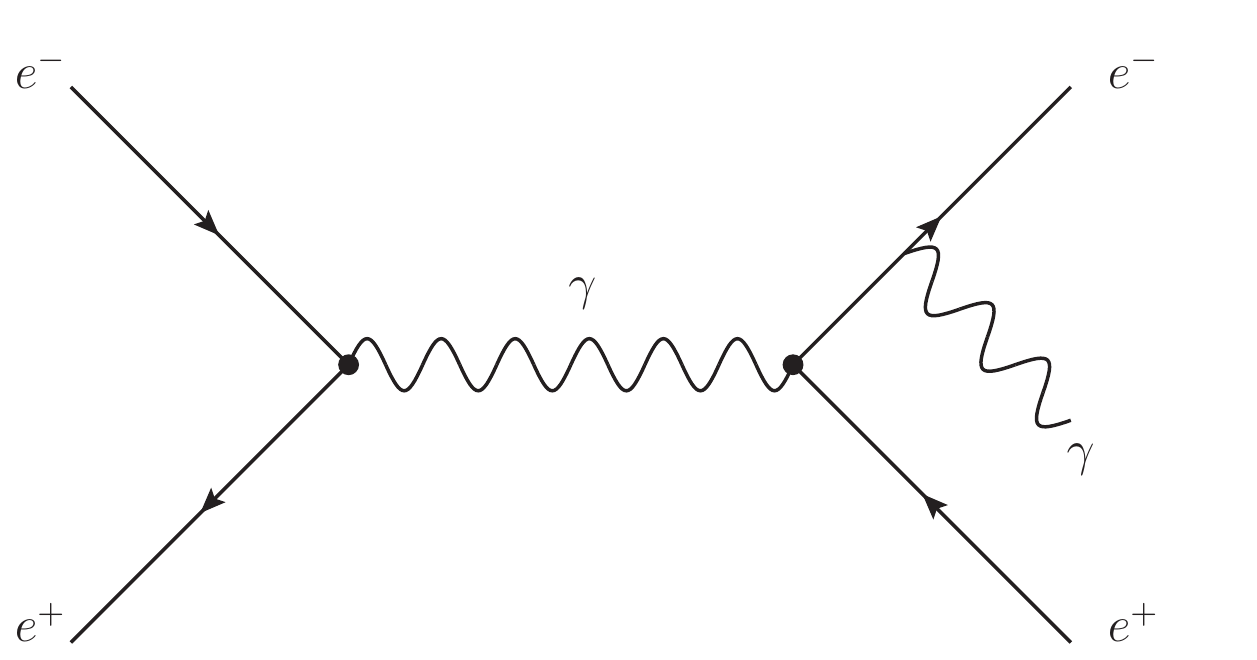}}
\end{center}
\caption[]{\emph{Interference of (a) and (b) is a sample contribution to $\sigma^{\rm{\ssN\ssL\ssO}}_{\gamma}$ in 
Eq.~(\ref{all-fermions}).}}
\label{AIRsoft}
\end{figure}

As a result, the following contributions will be studied:
\begin{itemize}
 \item
the $\sigma^{\rm{\ssN\ssN\ssL\ssO}}_{\mathit virt}$ consists of virtual \emph{two-loop corrections}
$ \sigma^{\rm{\ssN\ssN\ssL\ssO}}_{\mathit 2L}$
shown in Fig.~\ref{Ca}  and \emph{loop-by-loop corrections}
$ \sigma^{\rm{\ssN\ssN\ssL\ssO}}_{\mathit{1L1L}}$
shown in Fig.~\ref{Cb}:
\bea\label{eqa1}
\sigma^{\rm{\ssN\ssN\ssL\ssO}}_{\mathit virt} = \sigma^{\rm{\ssN\ssN\ssL\ssO}}_{\mathit 2L}
+
 \sigma^{\rm{\ssN\ssN\ssL\ssO}}_{\mathit{1L1L}}
\eea
\item
contributions with \emph{real photon emission}, shown in Fig.~\ref{AIRsoft}:
\bea\label{eqa111}
\sigma^{\rm{\ssN\ssL\ssO}}_{\gamma} =  \sigma^{\rm{\ssN\ssL\ssO}}_{\gamma,\mathit{soft}}(\omega) +
\sigma^{\rm{\ssN\ssL\ssO}}_{\gamma,\mathit{hard}}(\omega)
\eea
\item
contributions with \emph{real pair or hadron emission} depend a bit more on the flavour, as shown in 
Figs.~\ref{fig-36diagrams}-\ref{pions}:
\bea\label{eqa11}
\sigma^{\rm{\ssL\ssO}}_{\mathit real}
=
\sigma^{\rm{\ssL\ssO}}_{\mathit e^+e^-(e^+e^-)}
+
\sigma^{\rm{\ssL\ssO}}_{\mathit e^+e^-(f^+f^-)}
+
\sigma^{\rm{\ssL\ssO}}_{\mathit{e^+e^-(hadrons)}}
\eea
\end{itemize}
The self-energy blobs in Figs.~\ref{Ca}-\ref{AIRsoft} stand for lepton pair or hadronic self-energy insertions.

Technically, we have to re-order the corrections.
The following pieces will be summed up:
\begin{itemize}
 \item
virtual plus soft photonic corrections:
\bea\label{eq22}
\sigma^{\rm{\ssN\ssN\ssL\ssO}}_{v+s} = \sigma^{\rm{\ssN\ssN\ssL\ssO}}_{\mathit virt} +  \sigma^{\rm{\ssN\ssL\ssO}}_{\gamma,\mathit{soft}}(\omega),
\eea
see Figs.~\ref{Cb}-\ref{AIRsoft}.
The sum is infrared finite, but  depends on a soft-photon cut-off parameter $\omega$;
\item
hard photon radiation:
\bea\label{eq221}
\sigma^{\rm{\ssN\ssN\ssL\ssO}}_{h} = \sigma^{\rm{\ssN\ssL\ssO}}_{\gamma,\mathit{hard}}(\omega).
\eea
Here we take into account realistic experimental phase space cuts.
The sum $\sigma^{\rm{\ssN\ssN\ssL\ssO}}_{v+s} + \sigma^{\rm{\ssN\ssN\ssL\ssO}}_{h}$ now is independent of $\omega$.
\item
Bhabha scattering with the  additional production of fermionic pairs  or of hadrons,
$\sigma^{\rm{\ssL\ssO}}_{\mathit real}$.
\end{itemize}

\subsection{The virtual plus soft photon corrections $\sigma^{\rm{\ssN\ssN\ssL\ssO}}_{v+s}$\label{sec-virt-nnlo}}
The Feynman diagrams of the virtual corrections contain electron, heavy lepton or hadronic self-energy insertions.
We include in the cross section
{\it (i)} the irreducible vertex corrections of Fig.~\ref{Ca}(a) (see Eq. 39 in \cite{Actis:2008br});
{\it (ii)} 
the factorizable vertex and box corrections of Fig.~\ref{Ca}(b) and Fig.~\ref{Cb}(a,b) (see Eqs. 58, 60, 61  in \cite{Actis:2008br}); 
{\it (iii)} 
the irreducible box  corrections of Fig.~\ref{Ca}(c) (see Eq. 65 in
\cite{Actis:2008br}); 
{\it (iv)} the soft real photonic corrections (not shown, see Eqs. C4 and 62 in \cite{Actis:2008br}). 
For electrons, we will take the exact expressions, and for the other contributions the approximation $m_e^2<<s,|t|,|u|$ is applied.

The irreducible vertex diagrams are infrared finite, but the reducible vertices and the box contributions are not.
To make the latter two infrared finite, one has to add the corresponding soft real photon emission:
\begin{eqnarray}\label{fermions}
\frac{d \sigma^{\rm{\ssN\ssN\ssL\ssO}}_{v+s}} {d\Omega}
=
\frac{d \sigma^{\rm{\ssN\ssN\ssL\ssO}}_{\mathit{virt},e^+e^-}} {d\Omega}
+
\sum_{f =  \mu,\tau}
\frac{d \sigma^{\rm{\ssN\ssN\ssL\ssO}}_{\mathit{virt},f^+f^-}} {d\Omega}
+
 \frac{d \sigma^{\rm{\ssN\ssN\ssL\ssO}}_{\mathit{virt,had}}} {d\Omega}
+
\frac{d \sigma^{\rm{\ssN\ssL\ssO}}_{\mathit{virt}}} {d\Omega} \times
F_{\gamma,\mathit{soft}}(\omega)
\end{eqnarray}
Here, the term $\sigma^{\rm{\ssN\ssL\ssO}}_{\mathit{virt}}\times
F_{\gamma,\mathit{soft}}(\omega)$ arises from the interference of
(soft) single-photon bremsstrahlung diagrams, where one of the diagrams has a self-energy insertion.
The $\sigma^{\rm{\ssN\ssL\ssO}}_{\mathit{virt}}$ comes also from
a Bhabha  cross section  due to diagrams with first-order fermionic (hadronic) self-energy insertions, and $ F_{\gamma,\mathit{soft}}(\omega)$ is the usual soft photon eikonal factor; for explicit expressions see \cite{Actis:2007gi,Actis:2008br}.
The parameter $\omega$ is the infrared cut-off
\bea\label{eq-1}
\omega&=& E_{\gamma,\mathit{soft}}^{\max} ~~=~~ E_{\gamma,\mathit{hard}}^{\min}
\eea
 and has to be adapted such that soft-photon emission has  the Born kinematics and the sum of soft and hard photon radiation is numerically independent of $\omega$; typically, it is
$\omega/E_{\rm beam} = 10^{-6} \cdots  10^{-3}$.
The evaluation of the NNLO virtual corrections has been detailed elsewhere and we
may restrict ourselves here to few remarks on the specifics of this article.

In the simplest case of a one-loop self-energy insertion,
\begin{equation}
\label{PropReplace0}
\frac{g_{\mu\nu}}{q^2+i \delta}
\to
\Pi(q^2)~
\left(g_{\mu\nu} - \frac{q_\mu q_\nu}{q^2+i \delta}\right)
,
\end{equation}
one may just evaluate Feynman diagrams and gets after renormalisation for the case of a light fermion with  mass $m, m^2<<s$:
\bea\label{eq-11}
\Pi(s) &=& - \frac{\alpha}{3\pi}\left[\frac{5}{3} + \ln\left(-\frac{m^2}{s+i \delta}\right)\right].
\eea
For the virtual hadronic and  heavy lepton pair corrections we use the dispersion approach.
Here, the photon propagator is substituted in the Feynman diagrams as follows:
\begin{equation}
\label{PropReplace}
\frac{g_{\mu\nu}}{q^2+i \delta}
\to
\Pi(q^2)~
\left(g_{\mu\nu} - \frac{q_\mu q_\nu}{q^2+i \delta}\right)
~=~
 \frac{\alpha}{3 \pi}
 \int_{M_0^2}^{\infty}
 \frac{{ d }z~R(z)}{z}~
K_{\mathit SE}(q^2,z)
 \left(
 g_{\mu\nu} - \frac{q_\mu q_\nu}{q^2+i \delta}
 \right)
,
\end{equation}
with the propagator
\begin{equation}\label{eq-kernelSE}
K_{\mathit SE}(q^2,z) = \frac{1}{q^2-z+i \delta} .
\end{equation}
For heavy leptons with mass $m_l$ and charge $Q_l=-1$, it is at one loop (exact in the mass $m_l$):
\begin{eqnarray}\label{Rlept}
R_l(z; m_l)&=&
 Q_l^2\,\left(1+2  \frac{m_l^2}{z}\right)
        \sqrt{1-4 \frac{m_l^2}{z}}.
\end{eqnarray}

For  hadronic corrections, a natural choice for $R_{\mathit{had}}$ is a representation by experimental data:
\begin{eqnarray}
\label{Rhad}
R_{\mathit{had}}(z) &=&
\frac{\sigma_{\mathit had}(z)}
     {(4 \pi \alpha^2)\slash (3z)} \, ,
\end{eqnarray}
where $\sigma_{\mathit had}(z)\equiv \sigma(e^+e^-\to\gamma^\star\to \textrm{hadrons};z)$.
The real hadronic emission can be studied at present only for
pion pair production as only for this hadronic final state a MC code
is available.
Correspondingly, for the pion case we use instead of (\ref{Rhad})
the undressed pion form factor $F_\pi$:
\newcommand{\sigppm}{\sigma (e^+ e^- \to \pi^+ \pi^-) }
\newcommand{\eqc}{\;\;,}
\bea\label{sigpion}
\sigma_{\mathit had}(z) \to
\sigppm = \frac{\pi}{3}\frac{\alpha^2 \beta^3_\pi}{z}|F_\pi(z)|^2,
\eea
where
\bea\label{eq-12}
\beta_\pi=(1-4m_\pi^2/z)^{1/2} \, ,
\eea
 is the pion velocity.
For the cross section ratio $R$, which we need here,  this transforms to:
 \be
R_{\mathit{had}}(z) \to R_{\pi\pi}(z) = \frac{\beta^3_\pi}{4}|F_\pi (z)|^2.
\label{RPFF}
\ee
\color{black}
The pion form factor $F_{\pi}(z)$ has been determined in \cite{Bruch:2004py}. 
The numerical studies using this parameterisation are presented in Section \ref{sec-num}.

All this applies to one-loop insertions in reducible diagrams.
For irreducible two-loop vertex and box diagrams, one has to perform an additional loop integration, and the result is  more involved.
The complete virtual NNLO $N_f=1$  corrections (due to electron self-energy corrections) $\sigma^{\rm{\ssN\ssN\ssL\ssO}}_{v+s,e^+e^-} $ are known exact in $m_e$ \cite{Bonciani:2004gi} in terms of harmonic polylogarithms \cite{Remiddi:1999ew};
a  re-calculation  and also corresponding formulae in the kinematic limit $m_e^2<<s,t$, in terms of ordinary polylogarithms, are given in \cite{Actis:2007gi}.
The virtual NNLO  $N_f=2$  corrections (due to $\mu, \tau$, hadronic self-energy corrections)
are determined with dispersion formulas.

At the end of this Section, we shortly comment on
the structure of the various contributions to the virtual plus soft photon cross section:
\bea\label{eq-13}
\sigma^{\rm{\ssN\ssN\ssL\ssO}}_{v+s}
&=&
\sigma^{\rm{\ssN\ssN\ssL\ssO}}_{\mathit virt}
+  \sigma^{\rm{\ssN\ssL\ssO}}_{\gamma,\mathit{soft}}(\omega)
\nl
&=&
\sigma^{\rm{\ssN\ssN\ssL\ssO}}_{\mathit fact}
+
\sigma^{\rm{\ssN\ssN\ssL\ssO}}_{\mathit vert}
+
\sigma^{\rm{\ssN\ssN\ssL\ssO}}_{\mathit box}
+
\sigma^{\rm{\ssN\ssL\ssO}}_{\mathit{virt}} \times F_{\gamma,\mathit{soft}}(\omega).
\eea
The eikonal factor is in the limit of small $m_e$ \cite{Actis:2008br}:
\begin{eqnarray}\label{eq-18}
 F_{\gamma,\mathit{soft}}(\omega)
&=& \frac{\alpha}{\pi}
\Biggl\{
\left[
\frac{F_{\epsilon}}{\epsilon} -\ln\left(s/m_e^2\right) - 2\ln\left(\dfrac{2\omega}{\sqrt{s}} \right)
\right]
\left[-2\ln\left(s/m_e^2\right)+2 -2\ln\left( \frac{t}{u} \right) \right]
\\\nonumber
&&-~
\ln\left(s/m_e^2\right)^2 - 4\zeta_2
+ 2 \ln\left(s/m_e^2\right) +2\mathrm{Li}_2\left( -\frac{t}{u}\right) -  2 \mathrm{Li}_2\left( -\frac{u}{t}\right)
\Biggr\}.
\end{eqnarray}
The virtual contributions are:
\bea\label{eq-14}
\sigma^{\rm{\ssN\ssL\ssO}}_{\mathit virt}
 &\sim& \left( \frac{\alpha}{\pi} \right)^3
\left\{ \sigma^{\rm{\ssN\ssL\ssO}}_{{\mathit virt}, e^+e^-}
+
\Re \sum_{M} \int_{M_0^2}^{\infty} \frac{{ d }z~R_{M}(z)}{z}
\left[
C^{\rm{\ssN\ssL\ssO}}_{\mathit s}~K_{\mathit SE}(s,z) + C^{\rm{\ssN\ssL\ssO}}_{\mathit t} K_{\mathit SE}(t,z) \right]
\right\},
\\\label{eq-15}
\sigma^{\rm{\ssN\ssN\ssL\ssO}}_{\mathit fact}
 &\sim& \left( \frac{\alpha}{\pi} \right)^4
\left\{ \sigma^{\rm{\ssN\ssN\ssL\ssO}}_{{\mathit fact},e^+e^-}
+
\Re \sum_{M} \int_{M_0^2}^{\infty} \frac{{ d }z~R_{M}(z)}{z}
\left[
C^{\rm{\ssN\ssN\ssL\ssO}}_{\mathit fact,s}~K_{\mathit SE}(s,z) + C^{\rm{\ssN\ssN\ssL\ssO}}_{\mathit fact,t} K_{\mathit SE}(t,z) \right]
\right\} ,
\\ 
\label{eq-16}
\sigma^{\rm{\ssN\ssN\ssL\ssO}}_{\mathit vert}
&\sim&
\left( \frac{\alpha}{\pi} \right)^4
\left\{
\sigma^{\rm{\ssN\ssN\ssL\ssO}}_{\mathit vert,e^+e^-} +
\Re  \sum_{M} \int_{M_0^2}^{\infty} \frac{{ d }z~R_{M}(z)}{z}
\left[
C^{\rm{\ssN\ssN\ssL\ssO}}_{\mathit vert,s}~K_{\mathit vert}(s,z) + C^{\rm{\ssN\ssN\ssL\ssO}}_{\mathit{vert, t}}~ K_{\mathit vert}(t,z) \right]\right\},
\nl
\\ \label{eq-17}
\sigma^{\rm{\ssN\ssN\ssL\ssO}}_{\mathit box} &\sim& \left( \frac{\alpha}{\pi} \right)^4
\Biggl\{
\sigma^{\rm{\ssN\ssN\ssL\ssO}}_{\mathit box,e^+e^-} +
\Re  \sum_{M}\int_{M_0^2}^{\infty} \frac{{ d }z~R_{M}(z)}{z}
\Bigl\{
C^{\rm{\ssN\ssN\ssL\ssO}}_{\mathit{box,s}}
\bigl[ K_{\mathit box,A}(s,t,z) + K_{\mathit{ box,B}}(t,s,z)
\nonumber\\
&&+~K_{\mathit{box,C}}(u,t,z)
- K_{\mathit{ box,B}}(u,s,z)\bigr]
+ 
C^{\rm{\ssN\ssN\ssL\ssO}}_{\mathit{ box,t}}
\bigl[ K_{\mathit{box,B}}(s,t,z)
+ K_{\mathit{box,A}}(t,s,z)
\nonumber\\
&&
-~K_{\mathit{box,B}}(u,t,z)+ K_{\mathit{box,C}}(u,s,z)
\bigr] \Bigr\}\Biggr\}.
\eea
The sum over $M$ covers $m_{\mu},m_{\tau},m_{\pi}$ with the corresponding
parameterisations of $R_{M}(z)$; for leptons see \eqref{Rlept}.
The lower integration bound is $M_0^2 = 4m^2$ for leptons. For hadrons, where one sums over all the
hadronic contributions, the lower bound  is 
$M_0^2 = m_{\pi^0}^2$, corresponding to $\pi^0\gamma$, the lightest hadronic final state.
  
The kinematical factors $C$ (rational functions of $s$ and $t$) and the kernel functions $K$ are universal.
A special role play the  irreducible vertex  and box diagrams.
In these diagrams, the self-energy correction is part of a loop insertion, and due to the additional loop momentum integration the replacement (\ref{PropReplace})-(\ref{eq-kernelSE}) leads to more involved kernel functions compared to \eqref{eq-kernelSE}; for the vertex \cite{Kniehl:1988id}:
\begin{eqnarray}\label{eq:kernelV}
K_{\mathit vert}(x;z) = \frac{1}{3} \Bigl\{
 -   \frac{7}{8}
 -   \frac{z}{2 x}
 + \Bigl(  \frac{3}{4} + \frac{z}{2 x}  \Bigr)
   \ln\left(-\frac{x}{z}\right)
 -   \frac{1}{2}   \Bigl(  1 + \frac{z}{x}  \Bigr)^2
      \Bigl[  \zeta_2 - \text{Li}_2
      \left( 1 + \frac{x}{z} \right)  \Bigr]
\Bigr\}.
\end{eqnarray}
Here $\text{Li}_2(x)$ is the usual dilogarithm and $\zeta_2=\text{Li}_2(1) = \pi^2\slash 6$.
From the irreducible box diagrams we have three different, lengthy box kernel functions
$K_{\mathit{box,A}}(x,y,z), K_{\mathit{box,B}}(x,y,z), K_{\mathit{box,C}}(x,y,z)$;
see for explicit expressions Eqs. (71)-(73) of \cite{Actis:2008br}.

For practical reasons, it makes sense to split the $\sigma^{\rm{\ssN\ssN\ssL\ssO}}_{\mathit virt}$ into two pieces, namely the infrared finite  $\sigma^{\rm{\ssN\ssN\ssL\ssO}}_{\mathit vert}$ of \eqref{eq-16} and the so-called ``rest"  $\sigma^{\rm{\ssN\ssN\ssL\ssO}}_{\mathit rest}$,
\begin{eqnarray}\label{eq:v+rest}
\sigma^{\rm{\ssN\ssN\ssL\ssO}}_{\mathit virt} &=& 
\sigma^{\rm{\ssN\ssN\ssL\ssO}}_{\mathit vert} + \sigma^{\rm{\ssN\ssN\ssL\ssO}}_{\mathit rest}.
\end{eqnarray}
We just remind the reader that a third piece, the pure self-energy corrections, is not included in the study.
The $\sigma^{\rm{\ssN\ssN\ssL\ssO}}_{\mathit rest}$ is the sum of all the infrared divergent contributions. In~\cite{Actis:2008br}, it is detailed in Section VI 
and in Eq.~(87).
This sum is infrared finite, but depends on the photon energy cut $\omega$ related to the separation of soft and hard photons.
In the energy regions of relevance with $s>M_0^2$, the net result may adapted from Eq.~(93) of~\cite{Actis:2008br}:
\begin{eqnarray}\label{fromPRD-93}
\frac{ d{\sigma}_{\rm rest} }{ d\Omega }&=&
\frac{\alpha^4}{\pi^2 s}
\Biggl[
\int_{M_0^2}^{\infty}\, dz\, \frac{R(z)}{z}\,
\frac{1}{t-z}\, F_1(z)
\\
&&+~
\int_{M_0^2}^{\infty}\, dz\, \frac{1}{z\,\left(s-z\right)}\,
\Bigl\{
R(z)F_2(z)-R(s)F_2(s) + \left[ R(z)F_3(z)-R(s)F_3(s)\right] \ln\left| 1-\frac{z}{s}\right|
\Bigr\}
\nonumber\\
&&+~ \frac{R(s)}{s}\Bigl\{
F_2(s)\,\ln\Bigl(\frac{s}{M_0^2}-1\Bigr)
- 6\, \zeta_2\,F_4(s)
\nonumber\\
&&
+~F_3(s)\,\Bigl[
2\,\zeta_2
+~\frac{1}{2}\,\ln^2\Bigl(\frac{s}{M_0^2}-1\Bigr)
+\text{Li}_2\Bigl(1-\frac{s}{M_0^2}\Bigr)\,
\Bigr]
\Bigr\}
\Biggr].\nonumber
\end{eqnarray}
The explicit expressions for $F_1$ to $F_4$ are given in (88)-(91) of \cite{Actis:2008br}. They are infrared finite, but depend on $2\omega/\sqrt{s}$.

It is well-known that the irreducible vertex corrections from a fermion pair with mass $m$ contribute to the cross section with terms of order $\ln^3(s/m^2)$.
For electrons, this is a huge enhancement:
\bea\label{sig-irr-vert}
\sigma^{\rm{\ssN\ssN\ssL\ssO}}_{\mathit vert,e^+e^-}
&=&
C^{\rm{\ssN\ssN\ssL\ssO}}_{\mathit vert,e^+e^-,s} ~ \Re V_{2e}(s)
+
C^{\rm{\ssN\ssN\ssL\ssO}}_{\mathit vert,e^+e^-,t} ~  V_{2e}(t) ,
\eea
and the form factor is for $m_e^2/x<<1$ \cite{Burgers:1985qg}:
\bea \label{eq:V2}
V_{2e}(x)\!= \!\frac{1}{36} \ln^3\Bigl(-\frac{m_e^2}{x}\Bigr)
+ \frac{19}{72}\, \ln^2\Bigl(-\frac{m_e^2}{x}\Bigr)
+
v_{2e} .
\eea
The vertex function may be found in \cite{Bonciani:2003ai} exact in $m_e$.
For heavy leptons, the logarithmic terms  in $V_{2f}(x)$ agree with $V_{2e}(x)$, but a deviation appears in the constant term which becomes $v_{2f}$ \cite{Burgers:1985qg,Actis:2007gi}:
\bea\label{eq-189}
v_{2e}
&=&
\frac{1}{6}\,\left( \frac{265}{36}+\zeta_2 \right)\, \ln\Bigl(-\frac{m_e^2}{x}\Bigr) +~
\frac{1}{4}\left( \frac{383}{27} - \zeta_2 \right)+{\cal O}(m_e^2),
\\\label{eq-144}
v_{2f} &=&\frac{1}{6}\left( \frac{3355}{216} +\frac{19}{6}\,\zeta_2 - 2\,\zeta_3 \right)
+{\cal O}(m_f^2).
\eea
It is these logarithmic dependences which make the real pair emission contributions so important, because cancellations of the leading terms appear.

\bigskip

Technically, we evaluate the \emph{electron corrections} 
$d \sigma^{\rm{\ssN\ssN\ssL\ssO}}_{\mathit v+s,e} / d\Omega$ with the Mathematica program \textsc{CrossSection.m}
\cite{ACGR-bha-nnlo-ho:2011}.
The contribution 
$\frac{d \sigma^{\rm{\ssN\ssN\ssL\ssO}_{v+s,e}}} {d\Omega}$ is represented there by function
\textsc{NNLOel}. However, it includes also 
the iterated one-loop self-energies \textsc{NNLOfe2} and the 
genuine two-loop 
self-energy \textsc{NNLOfe1}.
Both are calculated separately, but are not included here in the numerics.
They have been subtracted from \textsc{NNLOel} in order to estimate what we call here  $\sigma^{\rm{\ssN\ssN\ssL\ssO}}_{\mathit virt,e}$.
The heavy fermion corrections have been calculated with a Fortran package applying the dispersion technique described here. 
In order to cover also pion pair corrections $\sigma_{v+s,\pi}^{\rm{\ssN\ssN\ssL\ssO}}$, the $\sigma_{v+s}^{\rm{\ssN\ssN\ssL\ssO}}$ is determined with an updated 
version \textsc{bha\_nnlo\_hf} of the Fortran package \textsc{bhbhnnlohf}~\cite{Actis:2008br,ACGR-bha-nnlo-ho:2011}.

\subsection{Hard photonic corrections $\sigma^{ \rm{ \ssN\ssN\ssL\ssO } }_{h}$ 
\label{sec-hard} }
The NNLO hard photonic corrections $\sigma^{\rm{\ssN\ssN\ssL\ssO}}_{h}$ with a self-energy insertion
 arise from the classes of diagrams shown in Fig.~\ref{AIRsoft}, with emission of one hard photon.
They were calculated
 with the Fortran program \textsc{Bhagen-1Ph-VAC} \cite{chunp} based
on the  generator  \textsc{Bhagen-1Ph} \cite{Caffo:1996mi}.
This cross section depends on the soft photon cut-off
$E_{\gamma}^{\min} = \omega$ and only after
adding them to $\sigma^{\rm{\ssN\ssN\ssL\ssO}}_{v+s}$, the sum of the
 two $\sigma^{\rm{\ssN\ssN\ssL\ssO}}_{v+s+h}$ is independent of the cut-off.
 We calculate here separately the contributions from diagrams with the
 electron, muon, tau, pion and complete hadronic vacuum polarisation insertions. The dependence on additional cuts is crucial and varies considerably with the experimental set-up.
A careful discussion is given in Section \ref{sec-num}.
 
Even if the insertion of the vacuum polarisation corrections to the
 square of the tree level
 amplitude \cite{Caffo:1996vi} is straightforward, for completeness we
 give below the
 formulae, which are used in the unpublished program \cite{chunp}.
We consider the process
\begin{equation}
\label{hardphoton}
  e^+ (p_+ ) + e^- (p_- ) \rightarrow e^+ (q_+ ) + e^- (q_- ) 
    +  \gamma (k) 
\end{equation}
and follow the notation of \cite{Caffo:1996vi}
\begin{equation}
\begin{array}{cccc}
     s  = (p_+ + p_-)^2  \ , & t  = (p_+ - q_+)^2\ , &u  = (p_+ -
     q_-)^2\ ,& s_1 = (q_+ + q_-)^2 ,
\\
     t_1 = (p_- - q_-)^2  \ ,& 
     u_1 = (p_- - q_+)^2  \ , &
     k_{\pm} = p_{\pm}.k  \ , &
     h_{\pm} = q_{\pm}.k  \ .  
\end{array}
\end{equation}
The differential cross section for the process (\ref{hardphoton}) can be written as
 \bea d\sigma  = {{\alpha^3}\over{2\pi^2 s}} (X+Y+Z) {{d^3q_+}\over{E_+}}
      {{d^3q_-}\over{E_-}} {{d^3k}\over{E_{\gamma}}}
       \delta^4(p_+ + p_- - q_+ - q_- - k) \ ,  \eea
where $E_+, E_-, E_{\gamma}$ are the energies of the final positron, 
electron and photon, respectively. 
The quantities $X, Y, Z$  refer to the $s$-channel annihilation, the $t$-channel scattering
 and the
interference part of the squared amplitude, respectively. Keeping only the
 diagrams with a virtual photon exchange (the original formulae \cite{Caffo:1996vi}
 contained also the weak $Z$ boson contributions) and with the
 vacuum polarisation corrections included they read:
\bea
\qquad \quad X &=&
    ({\rm Re}(\Pi(s)+\Pi(s_1)))(t^2 +t_1^2+u^2 +u_1^2 ) 
    {1\over{4 s s_1}}\left[{u\over{k_+h_-}} +{u_1\over{k_-h_+}} 
                         -{t\over{k_+h_+}} -{t_1\over{k_-h_-}} \right]
 \nonumber \\
 && +~ 2{\rm Re}(\Pi(s_1)) ((t^2 +t_1^2+u^2 +u_1^2 )
    {1\over{4 s_1 k_+  k_-}}   
  + 2{\rm Re}(\Pi(s))((t^2 +t_1^2+u^2 +u_1^2 ) 
    {1\over{4 s  h_+ h_-}}  \nonumber \\
 && -~2{\rm Re}(\Pi(s)){{m_e^2}\over{2 s^2}} \left[ 
      {{t_1^2}\over{(h_+)^2}} + {{t^2}\over{(h_-)^2}} 
   +  {{u^2}\over{(h_+)^2}} + {{u_1^2}\over{(h_-)^2}} 
   \right] \nonumber \\
 && -~2{\rm Re}(\Pi(s_1)){{m_e^2}\over{2 s_1^2}} \left[ 
     {{t_1^2}\over{(k_+)^2}} + {{t^2}\over{(k_-)^2}} 
   + {{u_1^2}\over{(k_+)^2}} + {{u^2}\over{(k_-)^2}} 
   \right] \ ,
\eea

\bea
Y &=&
    \left[(\Pi(t)+\Pi(t_1)) (s^2 +s_1^2+u^2 +u_1^2 )\right]
      {1\over{4 t t_1}} 
     \left[{u\over{k_+h_-}} + {u_1\over{k_-h_+}} + {s\over{k_+k_-}}
        + {s_1\over{h_+h_-}} \right]  \nonumber \\
  && -~2\Pi(t) (s^2 +s_1^2+u^2 +u_1^2 )
    {1\over{4 t k_- h_-}} 
   -2\Pi(t_1) (s^2 +s_1^2+u^2 +u_1^2 )
    {1\over{4 t_1 k_+ h_+}} \nonumber \\
  && -~2\Pi(t_1){{m_e^2}\over{2 t_1^2}} \left[ 
     {{s^2}\over{(h_+)^2}} + {{s_1^2}\over{(k_+)^2}} 
   + {{u^2}\over{(h_+)^2}} + {{u_1^2}\over{(k_+)^2}} 
     \right] \nonumber \\
  && -~2\Pi(t){{m_e^2}\over{2 t^2}} \left[ 
     {{s^2}\over{(h_-)^2}} + {{s_1^2}\over{(k_-)^2}}
   +  {{u_1^2}\over{(h_-)^2}} 
   +  {{u^2}\over{(k_-)^2}} 
     \right] \ , 
\eea
\bea
Z &=&
{{u^2 + u_1^2}\over 4} 
    \biggl[ 
{{\Pi(t)+{\rm Re}(\Pi(s))}\over{st}} 
     \left( {{u}\over {k_- h_+}} + {{s}\over{h_+h_-}} 
            - {{t}\over{k_-h_-}} \right) 
\nonumber \\
&&+~
   {{\Pi(t_1)+{\rm Re}(\Pi(s))}\over{st_1}} 
     \left( {{u_1}\over {k_+ h_-}} + {{s}\over{h_+h_-}} 
            - {{t_1}\over{k_+h_+}} \right) 
\nonumber \\
       &&    +~{{\Pi(t)+{\rm Re}(\Pi(s_1))  }\over{s_1t}} 
     \left( {{u_1}\over {k_+ h_-}} + {{s_1}\over{k_+k_-}} 
            - {{t}\over{k_-h_-}} \right) 
\nonumber \\
&& +~
   {{\Pi(t_1)+{\rm Re}(\Pi(s_1))}\over{s_1t_1}} 
     \left( {{u}\over {k_- h_+}} + {{s_1}\over{k_+k_-}} 
            - {{t_1}\over{k_+h_+}} \right) 
\biggr] 
\nonumber \\
 &&  -~{{m_e^2}\over{ s t_1}}  \left(\Pi(t_1)+{\rm Re}(\Pi(s))\right) {{u^2}\over{(h_+)^2}} 
    -{{m_e^2}\over{ s_1 t_1}} \left(\Pi(t_1)+{\rm Re}(\Pi(s_1))\right) {{u_1^2}\over{(k_+)^2}}  
\nonumber \\
   &&-~{{m_e^2}\over{ s t}}  \left(\Pi(t)+{\rm Re}(\Pi(s))\right) {{u_1^2}\over{(h_-)^2}} 
    -{{m_e^2}\over{ s_1 t}} \left(\Pi(t)+{\rm Re}(\Pi(s_1))\right) {{u^2}\over{(k_-)^2}} \ . 
 \eea 
The generation of the phase space was not changed with respect to the
 original program \textsc{Bhagen-1Ph} \cite{Caffo:1996mi}.

\subsection{Real electron pair contributions $\sigma^{\rm\ssL\ssO}_{e^+e^-(e^+e^-)}$ \label{sec-e}}
\begin{figure}[t]
\begin{center}
\subfloat[][]{\includegraphics[width=.3\textwidth]{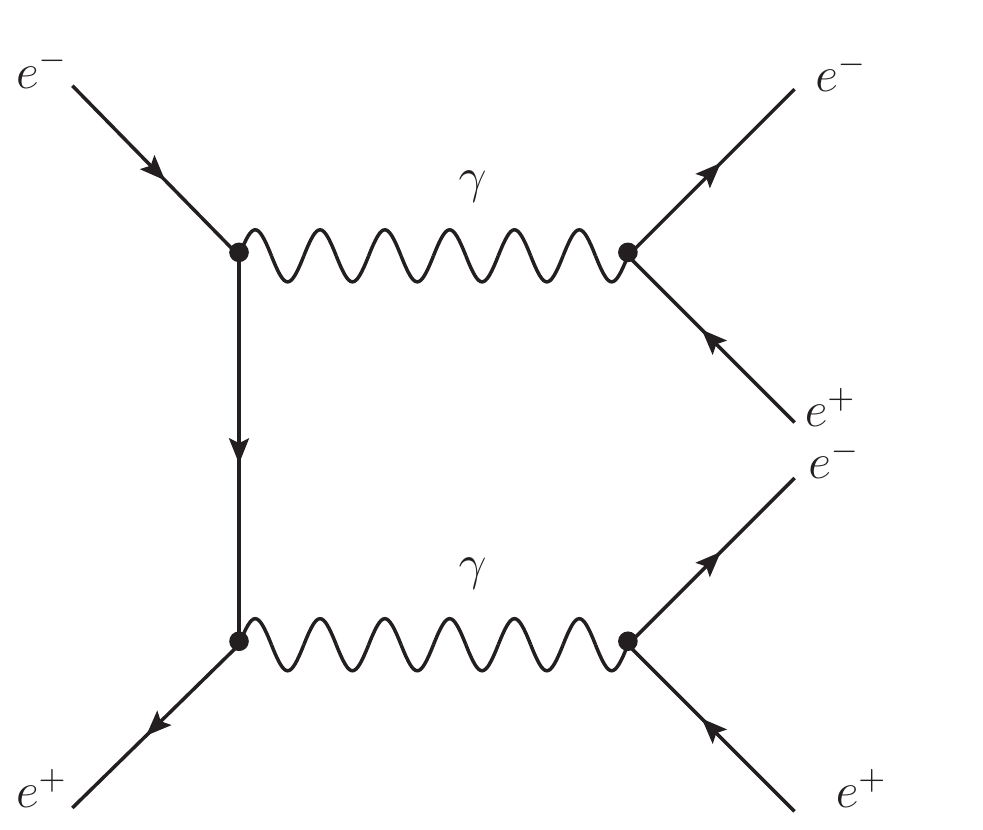}}
\subfloat[][]{\includegraphics[width=.3\textwidth]{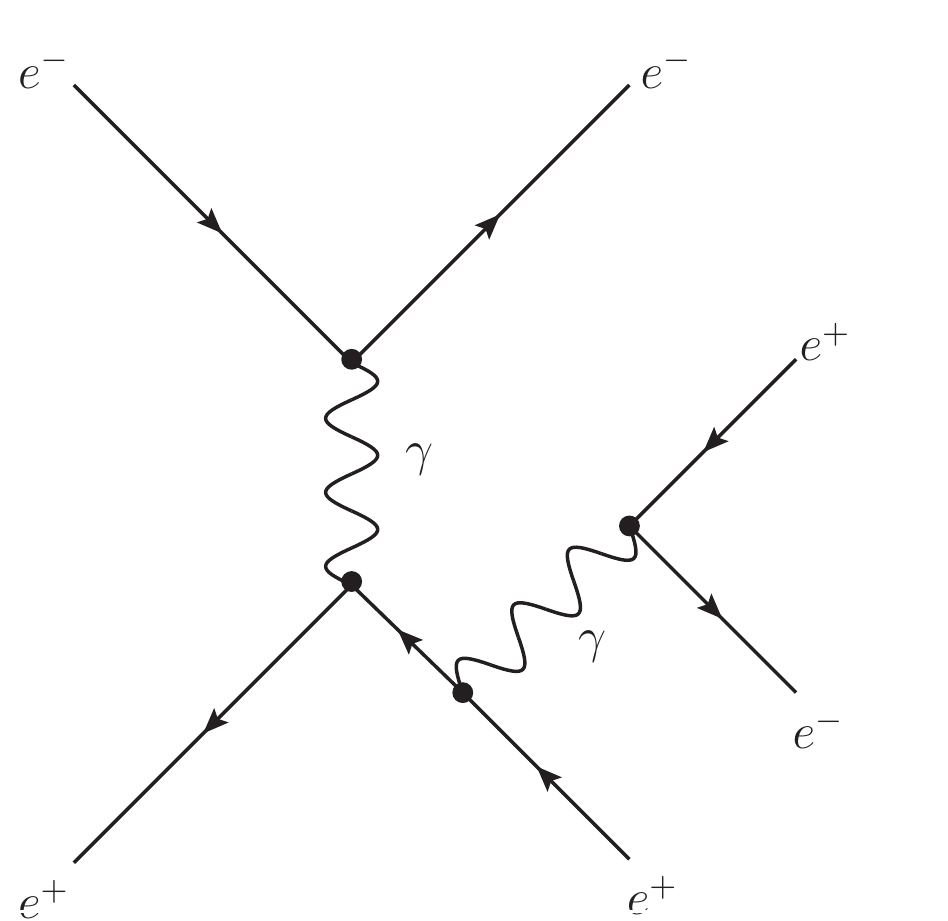}}
\subfloat[][]{\includegraphics[width=.18  \textwidth]{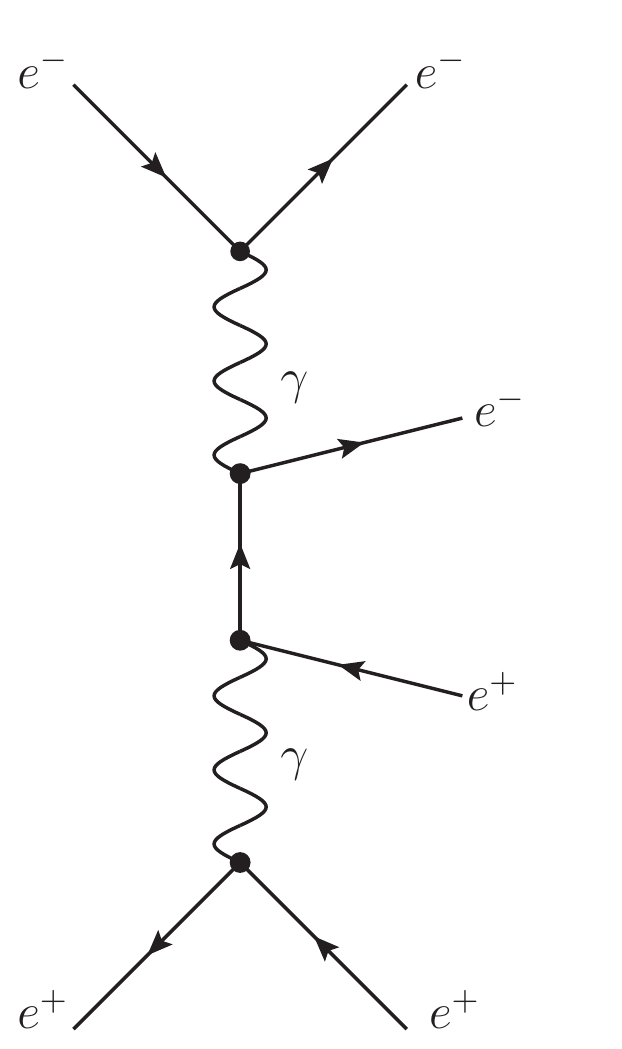}}
\end{center}
\caption[]{\emph{Samples of the 36 diagrams contributing to $e^+e^-\to e^+e^-e^+e^-$.}}
\label{fig-36diagrams}
\end{figure}

The most important real fermion pair corrections to Bhabha scattering are, at all energies, the unresolved electron pair corrections.
There are 36 diagrams of this kind contributing to
$\sigma^{\rm\ssL\ssO}_{e^+e^-(e^+e^-)}$, part of
the Bhabha cross section (\ref{all-fermions}).
Sample diagrams are shown in Fig.~\ref{fig-36diagrams}.

The $e^+e^-$ pair corrections fall into three classes:
four $s$-channel diagrams with two $e^+e^-$ pairs in the final state,
24 diagrams (8 in  $s$-channel and 16 in $t$-channel) with one $e^+e^-$ pair, and 8 peripheral $t$-channel diagrams (no  $e^+e^-$ pair).
What is usually considered as the electron pair corrections, are those with two electron pairs in the final state, Fig.~\ref{fig-36diagrams}(a).
The contribution from such soft electron pairs is known \cite{Arbuzov:1995vj}, see also \cite{Actis:2008br}.
It is, in the limit of small $m_e$ and small energy cut-off parameter $D$ of the unresolved $e^+e^-$ pair, proportional to the lowest order Bhabha 
Born cross section
$\sigma^{\rm\ssL\ssO}_{e^+e^-}$:
\begin{eqnarray}\label{sp}
\sigma^{\rm\ssL\ssO}_{e^+e^-(e^+e^-)}
&\sim&
 \sigma^{\rm\ssL\ssO}_{e^+e^-} ~\left(\frac{\alpha}{\pi}\right)^2~\delta^{e}_{\mathit soft},
\end{eqnarray}
with:
\begin{eqnarray}
\label{spe}
 \delta_{\mathit soft}^{e}&=& \frac{1}{3}\left[
\frac{1}{3}L_s^3+L_s^2\left(2\ln(D)-\frac{5}{3} \right) +L_s\left(4\ln^2(D)- \frac{20}{3}\ln(D)+A_s\right)
\right. \\ \nonumber &&
+~\frac{1}{3}L_t^3+L_t^2\left(2\ln(D)-\frac{5}{3} \right) +L_t\left(4\ln^2(D)- \frac{20}{3}\ln(D)+A_t\right)
\\ \nonumber &&
\left.
-~\frac{1}{3}L_u^3-L_u^2\left(2\ln(D)-\frac{5}{3} \right) -L_u\left(4\ln^2(D)- \frac{20}{3}\ln(D)+A_u\right)
 \right] ,
\end{eqnarray}
where
\begin{eqnarray}\label{spe1}
 L_s &=&\ln\left(\frac{s}{m_e^2} \right),
\\\label{spe2}
 L_v &=&\ln\left(-\frac{v}{m_e^2} \right), ~~~~v=t,u,
\\ \label{spe3}
A_s&=& \frac{56}{9}-4\zeta_2,
\\\label{spe4}
A_v&=& A_s+2\litwo\left( \frac{1\pm\cos \theta}{2}\right) , ~~~~v=t,u.
\end{eqnarray}
The parameter $D$ has to fulfill:
\bea\label{eq-10}
2m_e << D E_{\rm beam} << E_{\rm beam}.
\eea
 From the sum of (\ref{sp}) and (\ref{sig-irr-vert}), the compensation of the leading mass singularities (contained here 
 in the $L_s^3, L_t^3, L_u^3$ terms) in the cross section becomes evident.

If there are unresolved contributions, which do not fulfill Born kinematics, or if the logarithms are not really big, or if the experimental accuracy is at the per mille level or better, a complete calculation is needed, and this is part of the present study.

The Feynman diagrams are finite as long as the electron mass is assumed to be finite, so that a 
straightforward Feynman diagram calculation of the $2 \to 4$ process can be performed without any true singularities.

\subsection{Real muon and tau pair contributions  $\sigma^{\rm\ssL\ssO}_{e^+e^-(l^+l^-)}$\label{sec-mt}}
\begin{figure}[t]
\begin{center}
\subfloat[][]{\includegraphics[width=.25\textwidth]{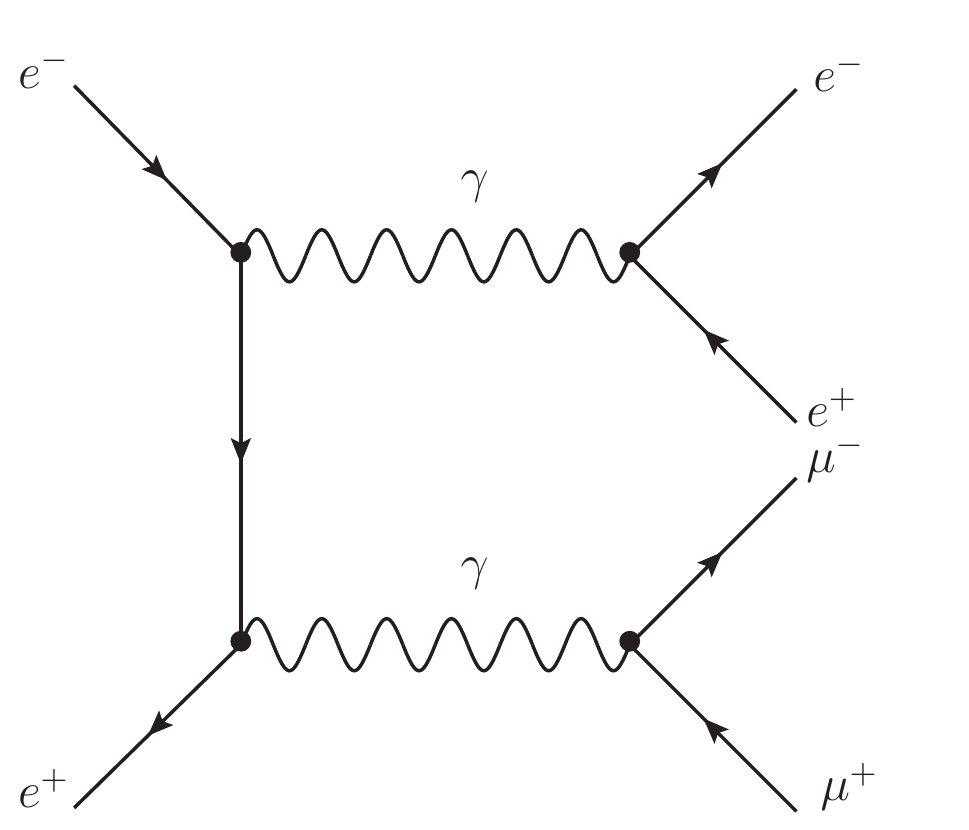}}
\subfloat[][]{\includegraphics[width=.28\textwidth]{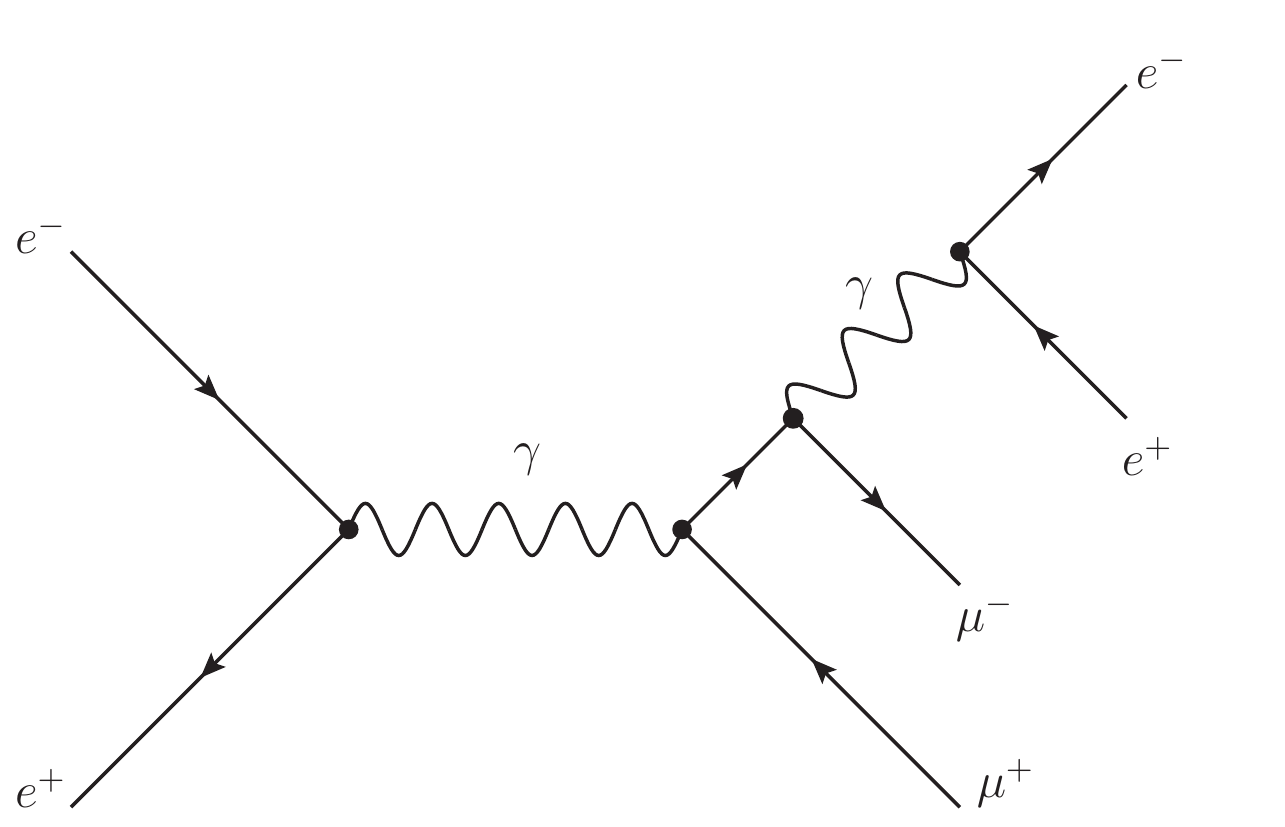}}
\subfloat[][]{\includegraphics[width=.18\textwidth]{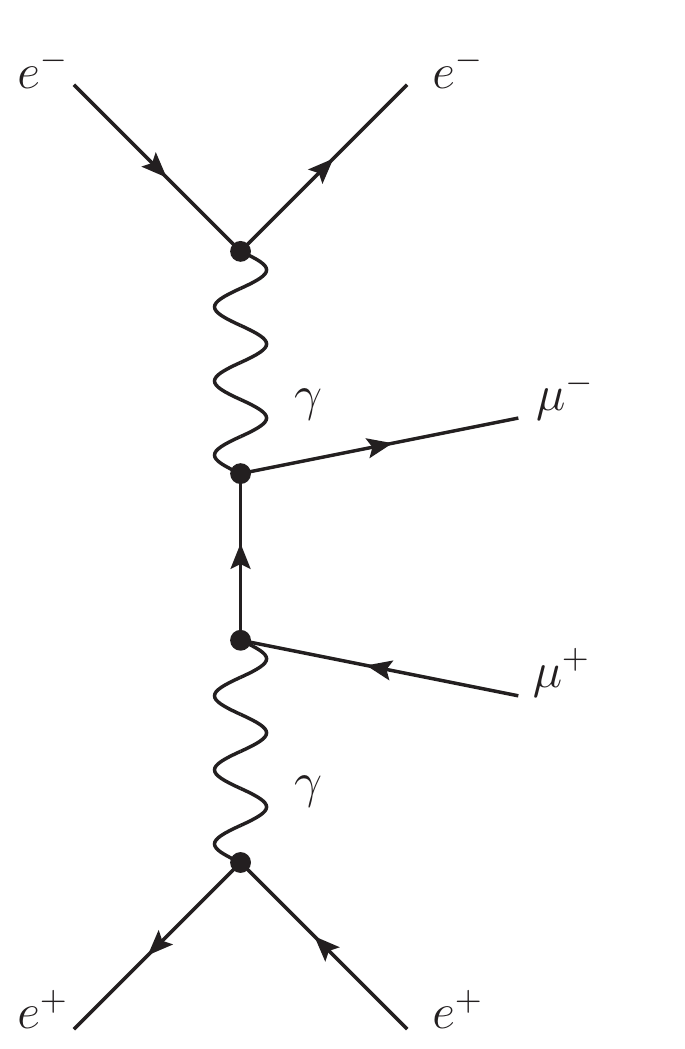}}
\subfloat[][]{\includegraphics[width=.28\textwidth]{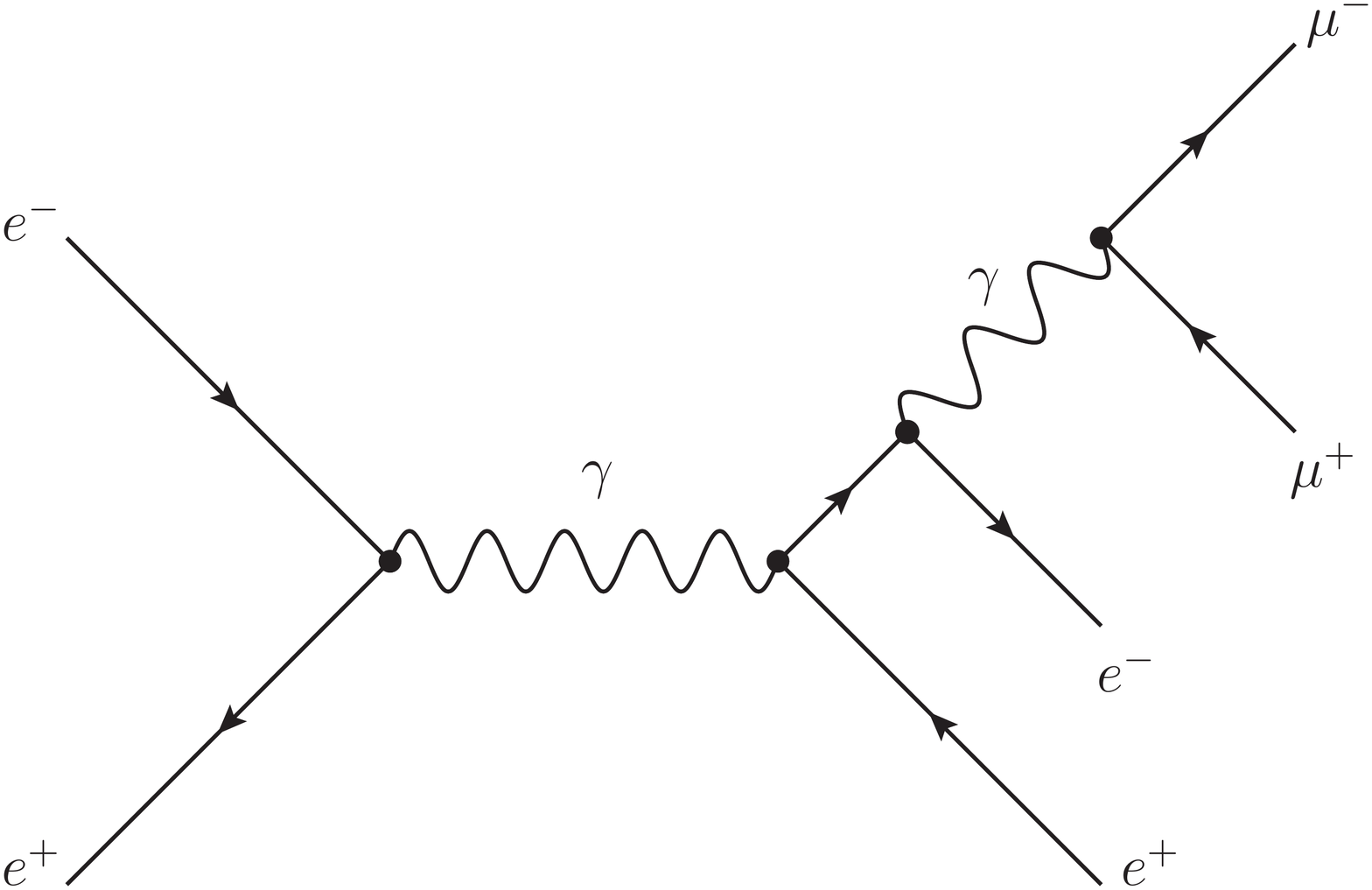}}
\end{center}
\caption[]{\emph{Samples of the 12 diagrams contributing to
 $e^+e^-\to e^+e^-\mu^+\mu^-$. A similar set of diagrams describes $e^+e^-\to e^+e^-\tau^+\tau^-$.}}
\label{fig-12diagrams}
\end{figure}

For both $e^+e^-\to e^+e^-\mu^+\mu^- $ and $e^+e^-\to e^+e^-\tau^+\tau^- $ there are 12 diagrams.
Samples of them  are shown in Fig.~\ref{fig-12diagrams}.

There are four classes of diagrams, to be discussed here for muon corrections:
two $s$-channel diagrams with
production of both an $e^+e^-$- and a $\mu^+\mu^-$-pair;
two $s$-channel diagrams with an $e^+e^-$-pair;
six diagrams (two in $s$-channel, four in $t$-channel) with
a fermion pair;
 and finally two peripheral $t$-channel diagrams.
Contrary to real electron  pair corrections, for heavy lepton pairs it is at the meson factory energies \emph{never} appropriate to assume that the lepton mass is much smaller than e.g. $\sqrt{s}$.
So, one has to perform a complete lowest order $2\to 4$ Feynman diagram calculation.
Again, the cross section is finite as long as all masses are assumed to
be finite.
The same discussion holds for the process
 $e^+e^-\to e^+e^-\tau^+\tau^- $.
\color{black}

The results for the electron, muon and tau pair corrections to the
 Bhabha scattering process  have been obtained in the
framework of the \textsc{Helac-Phegas} leading-order  MC program
\cite{Kanaki:2000ey,Cafarella:2007pc}.  
The phase space integration was executed with the help of
\textsc{Phegas} \cite{Papadopoulos:2000tt},  
a general purpose multi-channel  phase space
generator.  \textsc{Helac-Phegas} generates, in a fully automatic
manner, events for all possible parton level processes at hadron  and
lepton colliders within the Standard Model.  More precisely,
integrated cross sections and kinematic  distributions with
arbitrary cuts  on particles in the final state and   with full spin
correlations can be obtained.  It has  already been extensively used
and tested  in phenomenological studies, see e.g. 
\cite{Papadopoulos:2005ky,Gleisberg:2003bi,Alwall:2007fs,Englert:2008tn}.

In the present study, the exact QED $2 \to 4$ matrix elements for
$e^{+}e^{-}\to e^{+}e^{-}\ell^{+}\ell^{-}$ processes, where 
$\ell^{\pm}=e^{\pm},\mu^{\pm}, \tau^{\pm}$, have been generated, including
all Feynman diagrams (36, 12 and 12 respectively) and mass 
terms. Let us 
mention here, that in order to  generate pure QED contributions in 
\textsc{Helac-Phegas} the coupling of  $\ell^{+}\ell^{-}$ to the $Z$
boson has to be simply  set to zero. We have checked that the $Z$
 contributions are negligible for event selections used in this paper
  and do not affect any conclusions.

\subsection{Real pion pair contributions  $\sigma^{\rm\ssL\ssO}_{e^+e^-(\pi+\pi^-)}$\label{sec-p}}
\begin{figure}[t]
\begin{center}
\vspace*{2.0cm}
\subfloat[][]{\includegraphics[width=.3\textwidth]{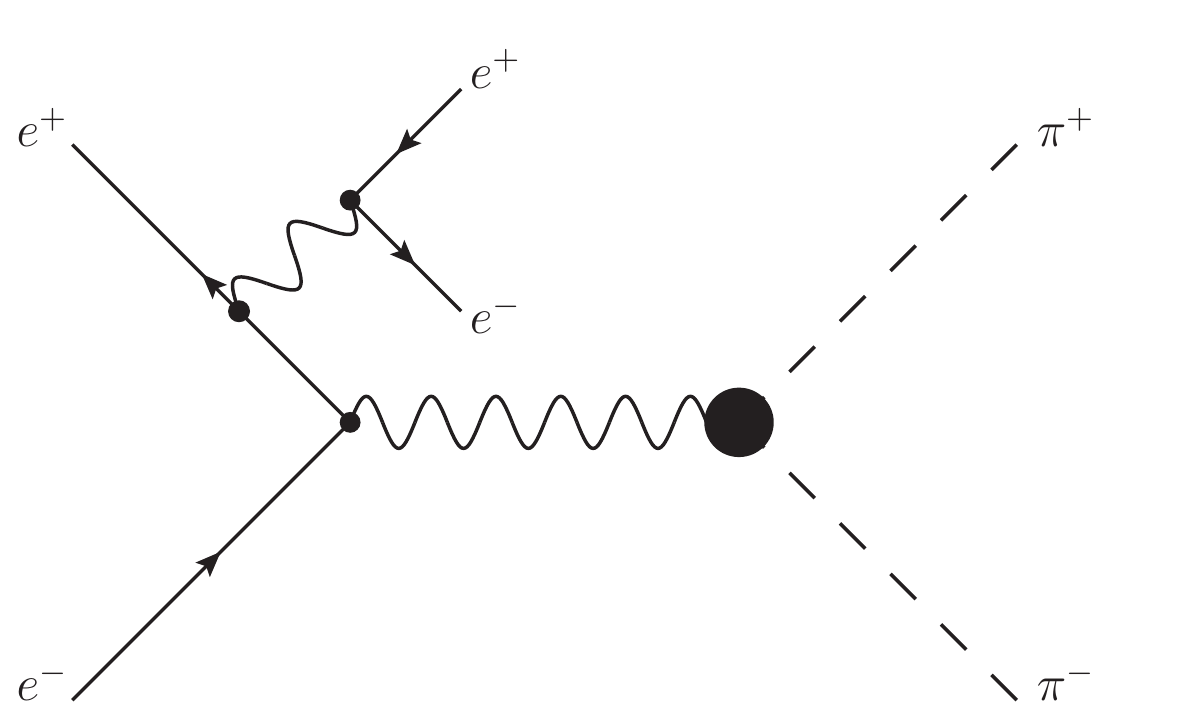}}
\subfloat[][]{\includegraphics[width=.3\textwidth]{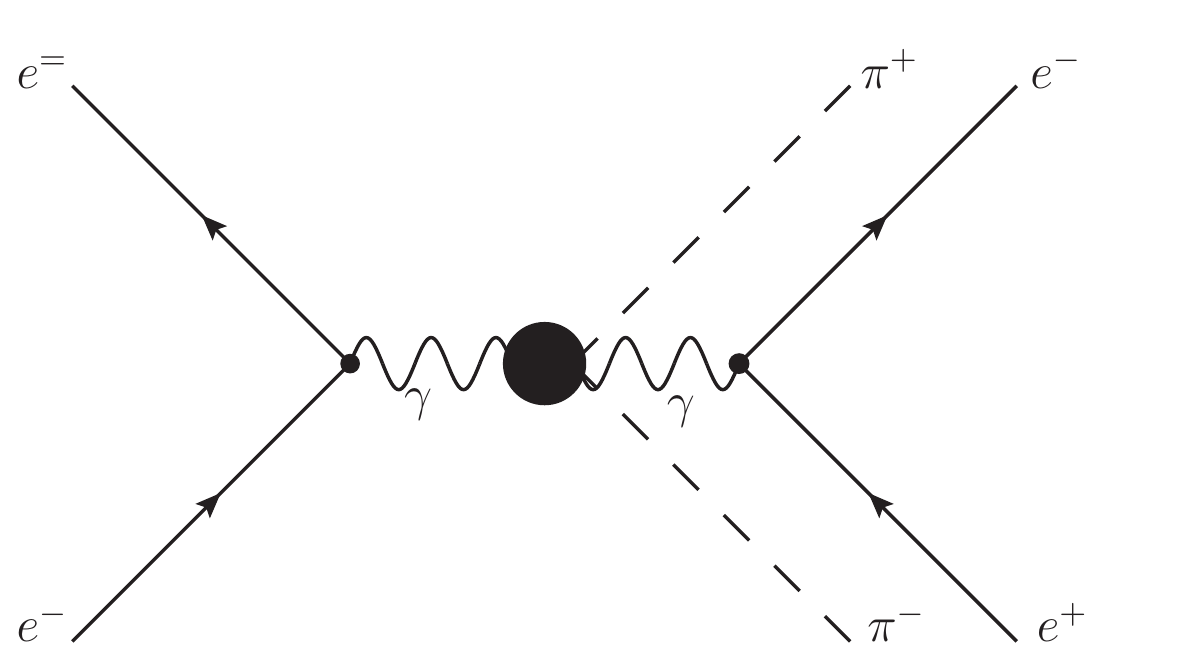}}
\subfloat[][]{\includegraphics[width=.3\textwidth]{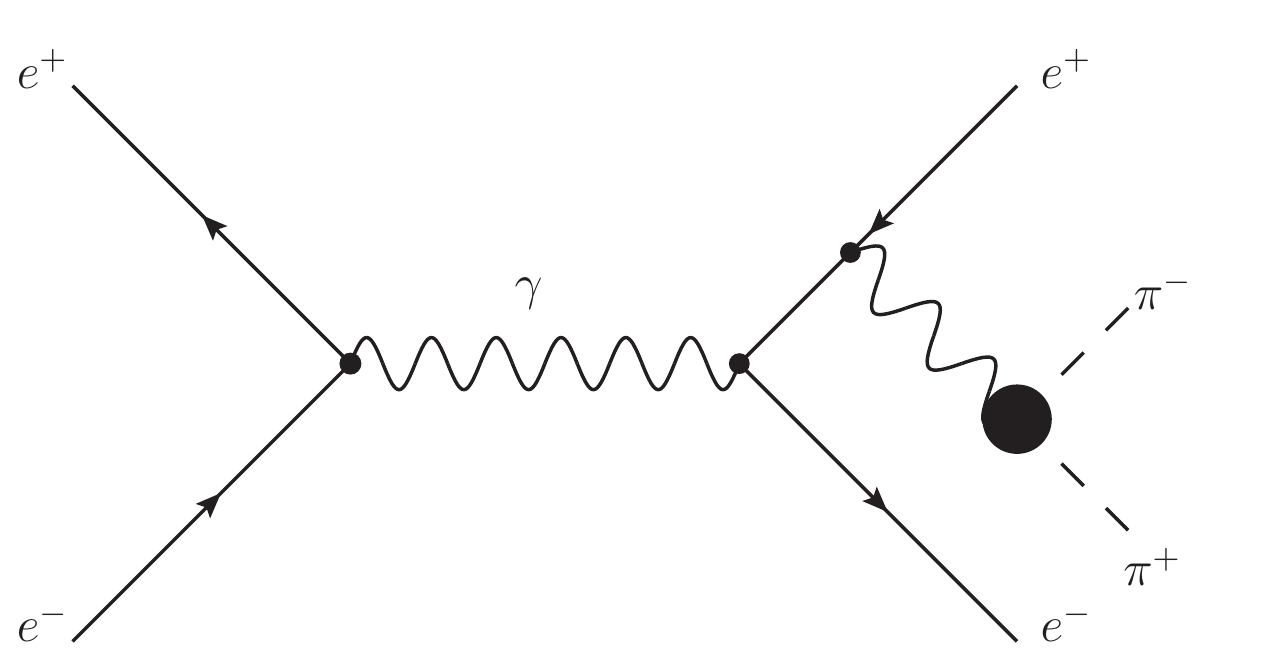}}
\\
\subfloat[][]{\includegraphics[width=.2\textwidth]{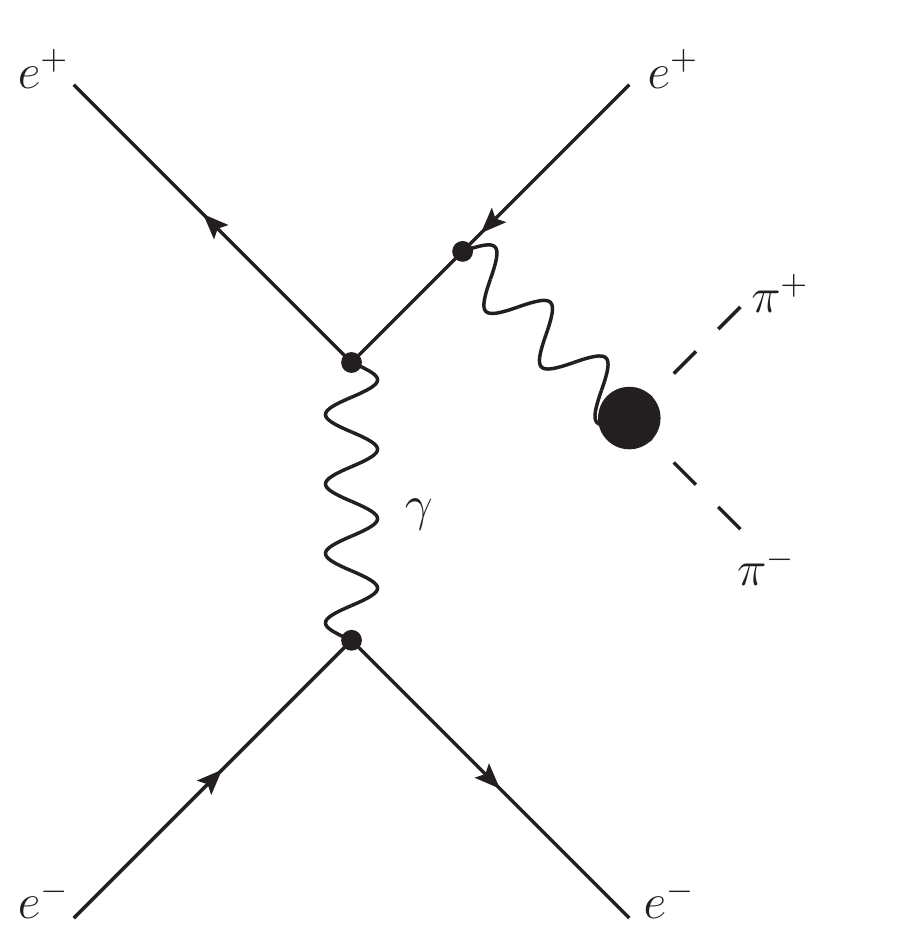}}
\subfloat[][]{\includegraphics[width=.2\textwidth]{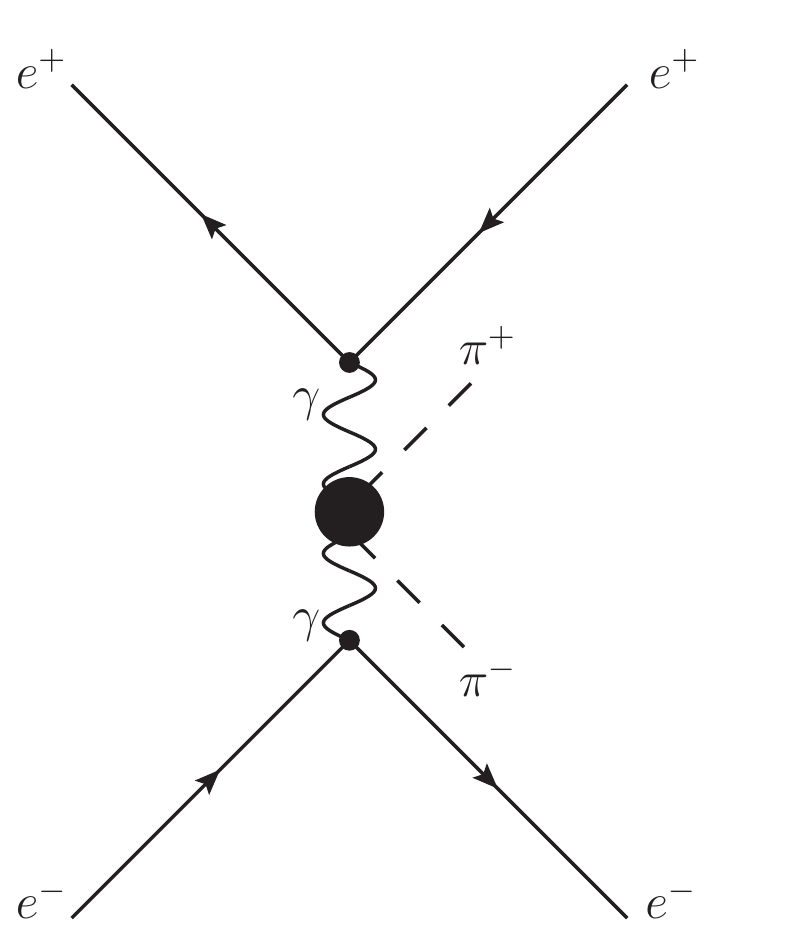}}
\end{center}
\caption[]{\emph{
Sample diagrams with real pion pair emission. 
}}
\label{pions}
\end{figure}

 The lightest hadronic final state produced in $e^+e^-$ scattering via
 the one photon exchange mechanism is the charged pion pair. This final
 state (i.e. $e^+e^-\pi^+\pi^-$) was investigated in details in
\cite{Czyz:2006dm,Czyz:2005ab,Czyz:2003gb} and implemented into the Monte
 Carlo generator \textsc{Ekhara} \cite{Czyz:2006dm,Czyz:2010sp}. Since other hadronic final states produced this
 way were never implemented in  a generator, their studies are
 impossible at present and results obtained for the reaction  $e^+e^- \to e^+e^-\pi^+\pi^-$
 will be used also as a hint towards understanding the importance of
 other reactions like  $e^+e^- \to e^+e^- +  (\pi^+\pi^-\pi^0, ~K^+K^-, ~K_S
 K_L, ~\cdots)$.
 The corresponding set of diagrams consists of 14 diagrams, with their
 representatives shown in Fig.~\ref{pions}. To model the pion--photon
 interactions we use the vector dominance model with the  pion form
 factor from \cite{Bruch:2004py}. These contributions can be seen as:
 initial state electron pair emission (a), final state electron/pion pair
 emission (b,c),
  pion pair emission from the $t$-channel Bhabha process
(d) and $\gamma^*-\gamma^*$ pion pair production (e).
 The last set of diagrams is small for large electron and positron
 angles and its modelling is relatively crude, neglecting scalar meson production
 and the subsequent decays to pion pairs.

\subsection{Other hadronic corrections  \label{sec-hadrons}}
As discussed in the previous Section the lack of an event
 generator for processes $e^+e^- \to e^+e^-$ plus hadrons does not allow
 the computation of the real hadron emission with the exception of
 charged pion pairs. However contributions coming from virtual hadronic vacuum
 polarisation insertions can be calculated completely 
(see Section \ref{sec-numNNLO}).
 In this Section we show how the vacuum polarisation insertions from pions
 compare to the full hadronic corrections. It has to be stressed that 
 even if this difference can give some indication of the size of
 the missing real
 emission contributions, its actual size depends heavily on the event
 selection used by a given experiment.
Thus a reliable
 estimation of the missing terms is not possible without Monte Carlo simulations.

Specific problems are caused by narrow resonances like $J/\psi, \psi(2S),...$
and one should devote them a special treatment.
Narrow resonances with mass $M_{\mathit res}$ and partial width $\Gamma^{e^+e^-}_{\mathit res}$ can be described approximately by the ansatz
\begin{equation}
R_{\mathit res}(z)= \frac{9 \pi}{\alpha^2} M_{\mathit res} \Gamma^{e^+ e^-}_{\mathit res}
\delta(z-M^2_{\mathit res})\, .
\label{eq-reson}
\end{equation}
Based on this, their contributions to the NNLO Bhabha process can be 
 derived from the general formulae of \cite{Actis:2008br}.
We discuss here as an example
 the contribution from the ``rest" (Eq.~\ref{fromPRD-93}); according to Eq. (87) of \cite{Actis:2008br} it reads: 
\begin{equation}
\frac{d {\sigma}_{\rm rest}}{d \Omega}=
\frac{9 \alpha^2}{\pi~s} ~\frac{\Gamma^{e^+ e^-}_{\rm res}}{M_{\rm res}}
\left\{
\frac{F_1(M^2_{\rm res})}{t-M^2_{\rm res}}
+ \frac{1}{s-M^2_{\rm res}} \left[ F_2(M_{\rm res}^2)
+ F_3(M_{\rm res}^2) \ln \left| 1-\frac{M_{\rm res}^2 }{s}\right|
\right]
\right\}.
\label{bhabha-res}
\end{equation} 
This is basically also Eq.~(E4) of \cite{Actis:2008br,ACGR-bha-nnlo-ho:2011}.
Eq. \ref{bhabha-res} becomes invalid when the center of mass energy comes too close to the position of a resonance, i.e. if $(s-M^2_{\rm res}) \lesssim \Gamma^{e^+ e^-}_{\rm res}M_{\rm res}$. In the numerical examples, Table~\ref{table-narrow-res}, this is not the case.

\begin{table}[h]
\setlength{\tabcolsep}{0.3pc}
\caption[]{\emph{
Soft+virtual NNLO contributions $\sigma_{\rm{rest,res}}^{\rm{\ssN\ssN\ssL\ssO}}$ from narrow resonances (n.r.)
defined by Eq.~(\ref{bhabha-res}) for the Bhabha
process  with
$\omega/E_{\rm beam}=10^{-4}$ (in nb).
The narrow resonance located closest to the center of mass energy of the given collider is
included (first column, ${\rm res}$) and excluded (second column, ${\rm res'}$). The third column contains the Born cross section.
}}
\label{table-narrow-res}
\begin{center}
\begin{tabular}{|lc|l|c|c|}
\hline
 & $\sqrt{s}$ &$\sigma_{\rm{rest,res}}^{\rm{\ssN\ssN\ssL\ssO}}$  & $\sigma_{\rm{rest,res'}}^{\rm{\ssN\ssN\ssL\ssO}}$ &  $\sigma_B$ 
\\
\hline
KLOE   & 1.020 & [all  n.r.]   &  [n.r. without J$\slash \psi$(1S)]
&       \\
       &       &  -0.04538  &   -0.0096      & 529.5  \\\hline
BES    & 3.097 & [all n.r.]    & [n.r. without J$\slash \psi$(1S)]
&        \\
       &       &  228.08         &   -0.0258        &  14.75  \\\hline
BES    & 3.650 &  [all n.r.]  & [n.r. without $ \psi$(2S)]     &
\\
       &       &   -0.1907        &    -0.023668        &  123.94  \\\hline
BES    & 3.686 &  [all n.r.]  & [n.r. without $ \psi$(2S)]       &
\\
       &       &   -62.537    &     -0.0254       &  121.53  \\\hline
BaBar & 10.56 &  [all n.r.]  & [n.r. without $\Upsilon$(4S)]      &
\\
       &       &  -0.0163       &  -0.01438          &  6.744 \\\hline
Belle  & 10.58 &  [all n.r.]  &[n.r. without $\Upsilon$(4S)]       &
\\
       &       &  0.04393   &    -0.0137        &    6.331  \\
\hline
\end{tabular}
\end{center}
\end{table}

To illustrate the role of narrow resonances, in Table~\ref{table-narrow-res} we show numerical results based on
Eq. (\ref{bhabha-res}).  We use parameters listed in Table~\ref{table:teub}. 
We can see that the contributions from narrow resonances dominate the NNLO Bhabha
correction for BES running at $J/\psi$ and $\psi(2S)$ energies.
For the remaining cases narrow resonances contribute below the per mille level
when compared to the Born cross section $\sigma_B$ or to
BabaYaga@NLO best predictions $\sigma_{\rm{\ssB\ssY}}$, see Table~\ref{table-compar-NNLO-net}.

We conclude that for experiments performed on top of a narrow
resonance, this resonance cannot be treated as a mere correction and more detailed studies have to be performed.
  These should include examining of finite width effects, beam spread effects, estimation of NNNLO corrections
  and the accuracy of the vacuum polarisation insertions in a close
  vicinity of these resonances. Having this in mind we 
 do not present here hadronic contributions for the BES-III experiment running
 at $J/\psi$ and $\psi(2S)$ energies and we plan to devote to this issue a
 separate study. 

We now come to the net hadronic vacuum polarization effects, i.e. look now at the sum of the so-called ``rest" terms and the irreducible vertex corrections. 
To obtain all numerical results below
we use ``rest" as given by Eq. \ref{fromPRD-93} together  with the corresponding formula for the vertex, Eq. \eqref{eq-16}.
The recent update of $R_{had}$ valid in the range $ m_{\pi_0}^2< s< (100\;$ GeV$)^2$ \cite{teubner:2010} is applied, and  for higher
$s$ the $R_{had}$ is taken from \cite{Harlander:2002ur}. For further
details on the implementation of $R_{had}$ see Appendix E of \cite{Actis:2008br}. 

The numerical integrations for the hadronic virtual and soft contributions  were performed by means of the adaptive integration routine VEGAS \cite{vegas},
which works efficiently even for so narrow resonances like $J/\psi$.

\begin{table}[ht]
\caption[]{\emph{Parameters of narrow resonances used in 
~\cite{teubner:2010} (T. Teubner, private information).}}
\centering
\setlength{\arraycolsep}{\tabcolsep}
\renewcommand\arraystretch{1.1}
\begin{tabular}{|r|r|r|}
\hline
resonance & $M_{\rm res}$ [GeV] & $\Gamma^{e^+ e^-}_{\rm res}$ [keV] \\
\hline
\hline
J$\slash \psi$(1S) & 3.096916 & 5.55\\
$\psi$(2S) & 3.686093 & 2.33\\
$\Upsilon$(1S) & 9.46030 &1.34\\
$\Upsilon$(2S)& 10.02326 & 0.612 \\
$\Upsilon$(3S)& 10.3552 & 0.443\\
$\Upsilon$(4S)& 10.5794 & 0.272 \\
$\Upsilon$(5S)& 10.865 &  0.31\\
$\Upsilon$(6S)& 11.019 &  0.13\\
\hline
\end{tabular}
\label{table:teub}
\end{table}

\begin{figure}[t]
\begin{center}
\subfloat[][]{\includegraphics[width=.5\textwidth]{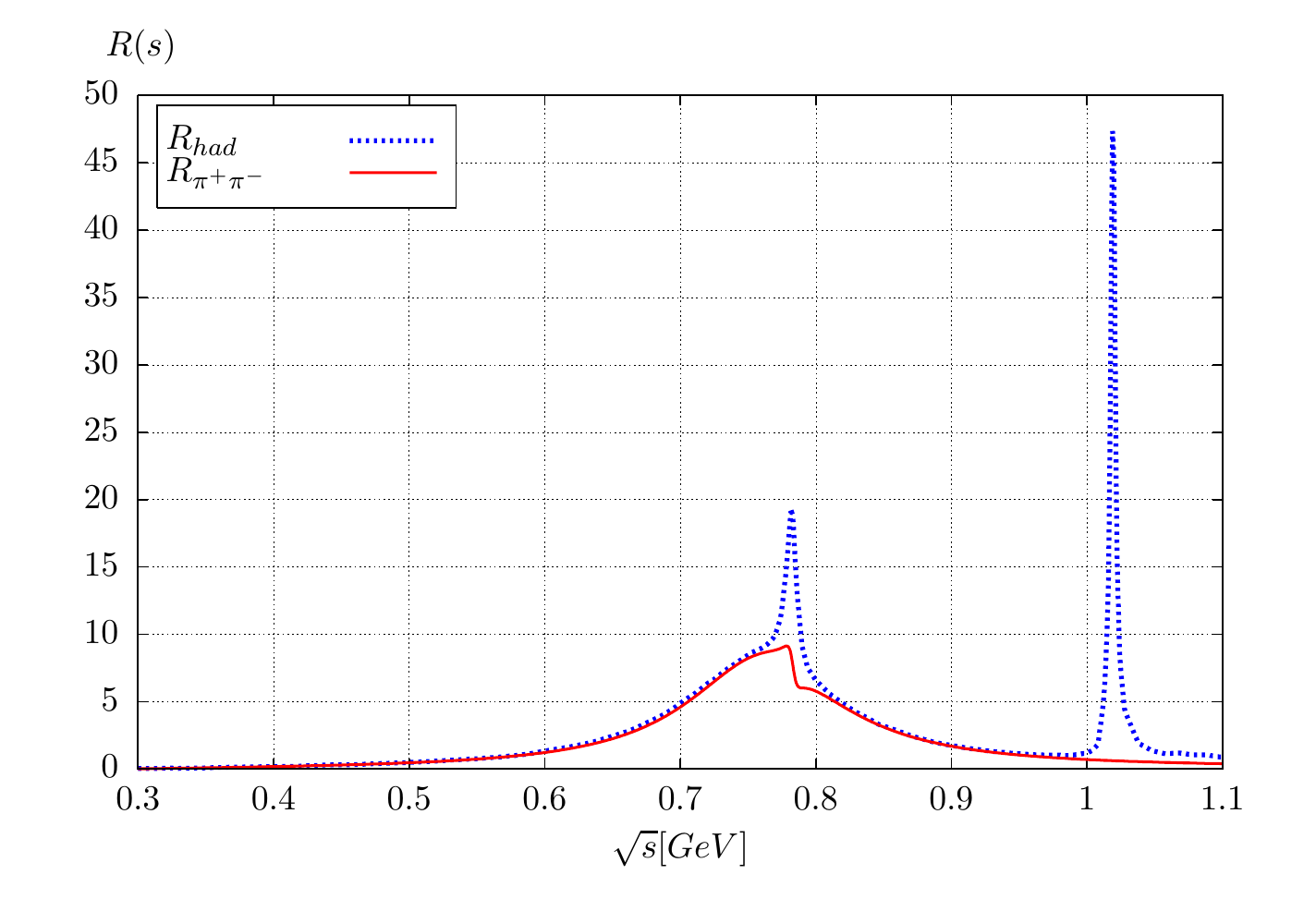}}
\subfloat[][]{\includegraphics[width=.5\textwidth]{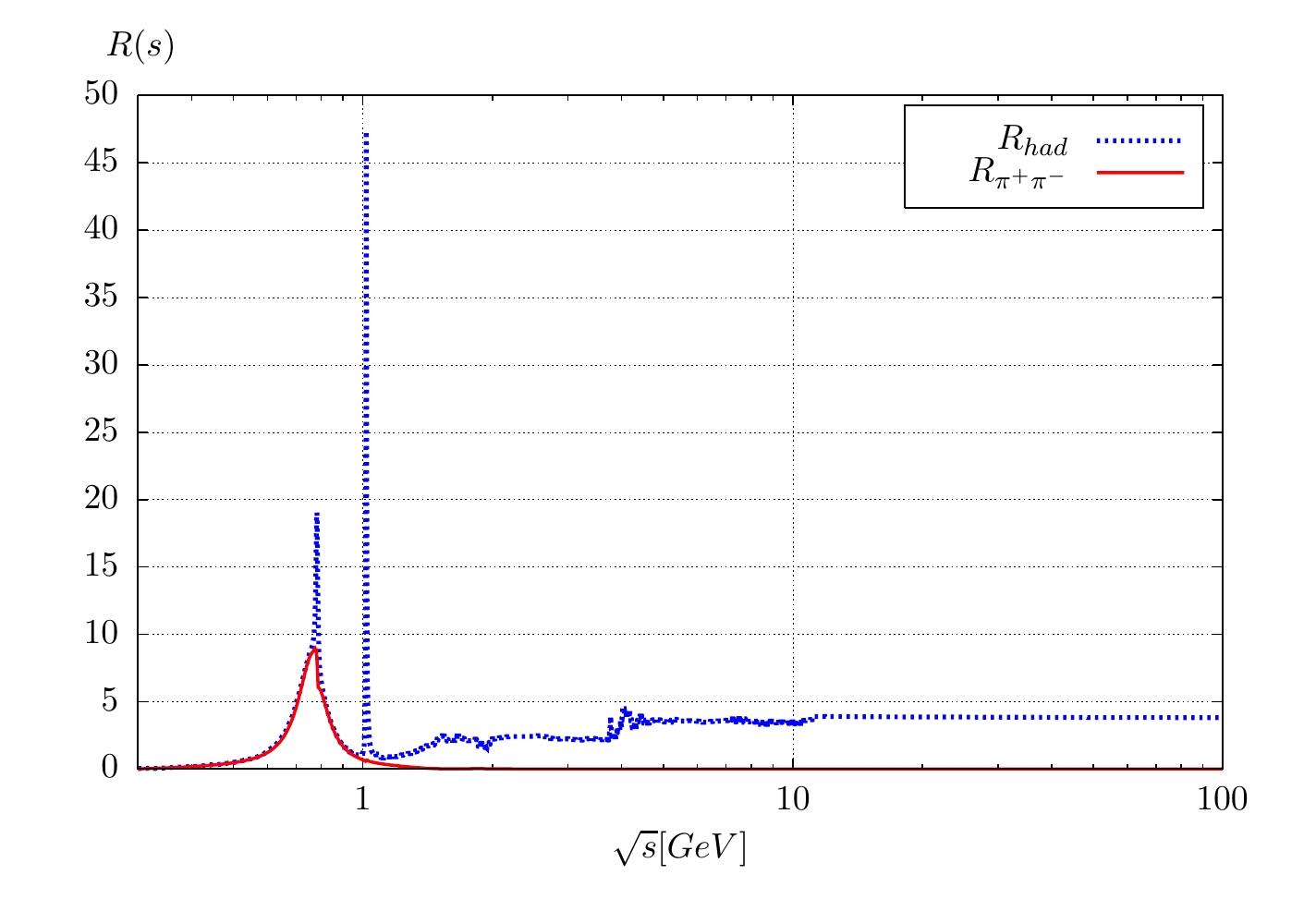}}
\end{center}
\caption{\emph{The $R_{\mathit{had}}$ vs. $R_{\mathit{\pi^+\pi^-}}$. 
Narrow resonances are not included.
\label{fig-had-res-nores}}} 
\end{figure}

In Fig.~\ref{fig-had-res-nores} we compare the full result for 
  $R_{had}$ and the contributions coming from pions only. At low energies 
 (Fig.~\ref{fig-had-res-nores} a) the biggest difference
  comes mostly from contributions of three pions and from kaon pairs
 pronounced at $\omega$  and $\phi$ resonances, while at high energies
 the pion contributions vanish rapidly and do not play any significant
  role (Fig.~\ref{fig-had-res-nores} b).
 
\begin{table}[t]
\caption{\emph{Comparison of hadronic contributions modelled by $R_{\pi^+\pi^-}$ and 
$R_{had}$. For hadrons, real emission is restricted to pions only.}
\label{tab-sma-pi-had}}
\centering
\setlength{\arraycolsep}{\tabcolsep}
\renewcommand\arraystretch{1.1}
\begin{tabular}{|r|r|r|r|}
\hline
 & KLOE & BES & BaBar  \\
\hline
\hline
$\sigma_{S+V}$,  $R_{\pi^+\pi^-}$ & -1.36  & -0.818 & -0.0533 \\
\hline
$\sigma_{S+V}$, $R_{had}$
& -1.06 & -1.81  &  -0.1888\\
\hline
$\sigma_{S+V+H}$, $R_{\pi^+\pi^-}$ & -0.186  & -0.0447  & -0.00229 \\
\hline
$\sigma_{S+V+H}$, $R_{had}$
& 0.47 & -0.15 & -0.0088 \\
\hline
\end{tabular}
\end{table}

From Table \ref{tab-sma-pi-had} it is clear that 
the pion contributions are not sufficient for an accurate evaluation
 of the hadronic contributions and the lack of a generator for generic
  hadronic final states does not allow to draw a final conclusion
  about the real hadronic corrections with the exception of the charged
  pion pair emission and event selections which kinematically exclude
  the hadron production (KLOE energy).

\begin{figure}[t]
\begin{center}
\includegraphics[scale=0.85]{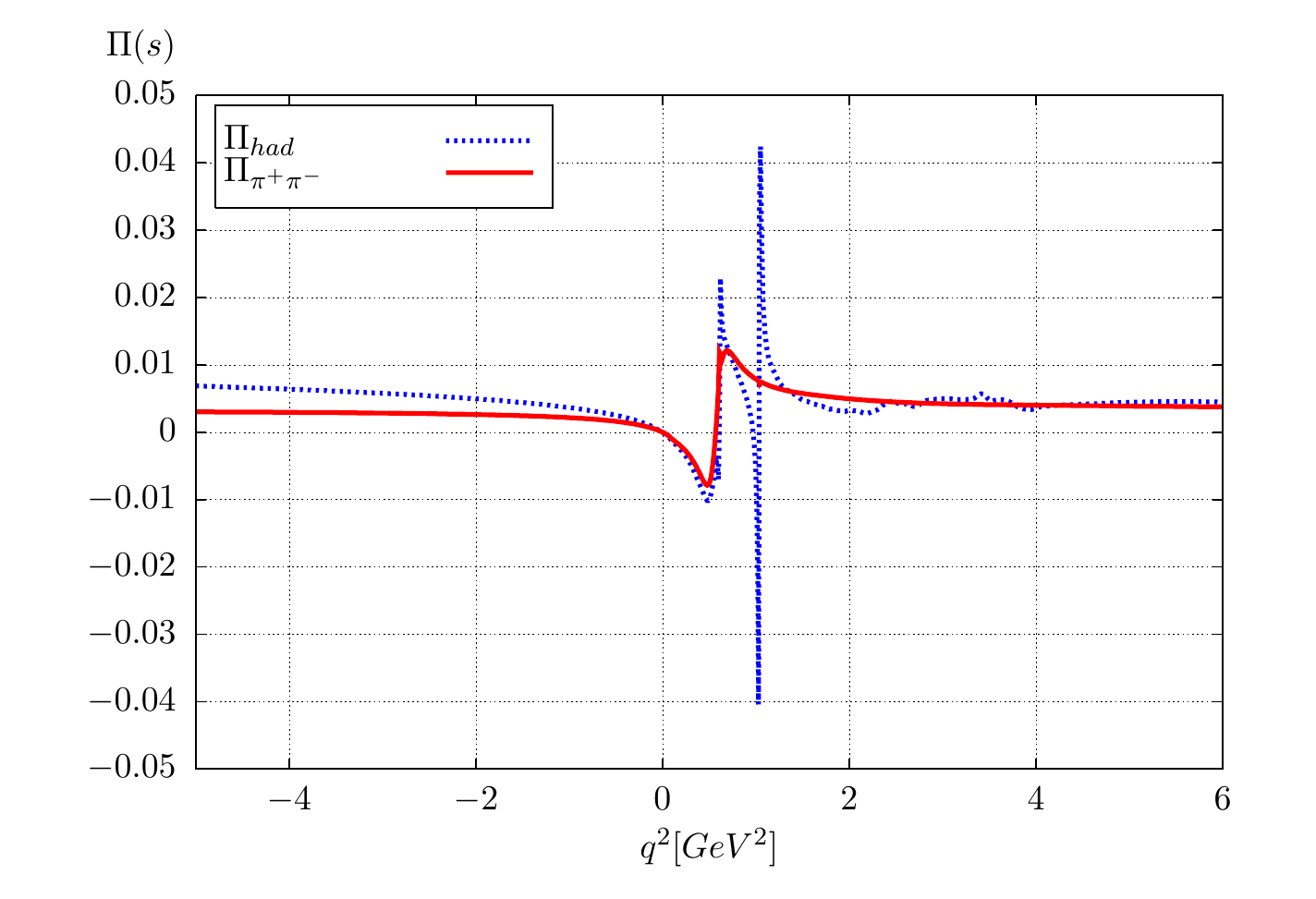}
\end{center}
\caption{\emph{Hadronic vs. pion pair contributions to the real part of the vacuum polarisation function. }
 }\label{intR}
\end{figure}

  The $R$ enters the results with  weight functions, which give
  more relevance to the low energy range, however the differences are 
  important also there. Partially one can see the effect of the  weight
   functions comparing the full hadronic corrections
  to the vacuum polarisation with pion pair contributions (Fig. \ref{intR}),
  however the complete weight functions are complicated and different
  for the virtual and real contributions, so the careful evaluation
  of the integrals is needed.

\subsection{Exact NNLO numerical results\label{sec-numNNLO}}

In this Section we collect exact NNLO results which can serve as a benchmark
for further investigations. The event selections used here are very close
to event selections used at meson factories to measure their luminosity and are described in
Appendix \ref{app-cuts}. We give separately the contributions 
from electron, muon, tau and pion pair production as well as the
complete hadronic contribution. For the last one, as mentioned already in the
previous Section, there exist no generator to give the contributions from
the real hadron emission beyond the pion pair production.  An educated
guess of the size of the missing contributions,
 based on the fact that the pions are the lightest hadrons produced
and that the highest energy of the meson factories is about 10 GeV,
is that they should not be much higher than the contributions from
 pion pairs. Thus based on information from Table {\ref{table-NNLO-pi}}
 we can conclude 
 that they should be completely negligible for the event selections 
  used at meson factories for the luminosity measurements.
 
In Tables \ref{table-el-NNLO}-\ref{table-NNLO-had} the meaning of the different entries is the
following: $\sigma_{B}$ is the Born cross section, $\sigma_{v+s}$ the cross
section with NNLO virtual plus soft photon corrections (Section \ref{sec-virt-nnlo}),  $\sigma_{h}$ the 
NNLO cross section with a self energy insertion corrected by the emission of one hard
photon (Section \ref{sec-hard}), $\sigma_{v+s+h}$ the sum of $\sigma_{v+s}$ and $\sigma_{h}$ and
$\sigma_{\rm pairs}$ the leading-order cross section with emission of real pairs (Sections \ref{sec-e}-\ref{sec-p}). 
The total NNLO massive correction can be obtained by summing $\sigma_{v+s+h}$ with $\sigma_{\rm pairs}$.

\begin{table}[h]
\setlength{\tabcolsep}{0.3pc}
\caption[]{\emph{%
Results for electron pair corrections at different energies, in GeV,
for the reference event selection defined in Appendix A.
The $\sigma_{B}$ is the Born cross section within the acceptance cuts.
The separation of soft and hard photons is fixed at
$\omega / E_{\rm beam}=10^{-4}$,
where $\omega = E_{\gamma}^{\min}$ (soft photon cut-off).
All cross sections are given in nb.
}}
\label{table-el-NNLO}
\begin{center}
\begin{tabular}{lcccccc}
\hline
$e^+e^-$& $\sqrt{s}$ &$\sigma_{B}$ & $\sigma_{h}$ & $\sigma_{v+s}$ &
 $\sigma_{v+s+h} $ & $\sigma_{\rm pairs}$
\\
\hline
KLOE   & 1.020 & 542.663(6)  & 9.5021(2)   & -11.5666    & -2.0645(2)   & 0.2712(15)  \\
BES  & 3.097 & 173.98(2)   & 4.16202(13) & -4.71708    & -0.55506(13) & 0.19977(116)  \\
BES  & 3.650 & 129.0958(4) & 3.19544(9)  & -3.55544    & -0.36000(9)  & 0.188856(997)  \\
BES  & 3.686 & 123.32(1)   & 3.14439(9)  & -3.49579    & -0.35140(9)  & 0.18740(99) \\
BaBar  & 10.56 & 5.481(1)    & 0.202439(7) & -0.223667   & -0.021228(7)  & 0.01355(8)     \\
Belle  & 10.58 & 6.73555(4)  & 0.21572(7)  & -0.25596    & -0.04024(7)   & 0.0130999(469)  \\
\hline
\end{tabular}
\end{center}
\end{table}

\begin{table}[h]
\setlength{\tabcolsep}{0.3pc}
\caption[]{\emph{
Results for muon pair corrections at different energies, in GeV,
for the reference event selection defined in Appendix A.
The $\sigma_{B}$ is the Born cross section within the acceptance cuts.
The separation of soft and hard photons is fixed at
$\omega / E_{\rm beam}=10^{-4}$,
where $\omega = E_{\gamma}^{\min}$ (soft photon cut-off).
All cross sections are given in nb.
}}
\label{table-mu-NNLO}
\begin{center}
\begin{tabular}{lcccccc}
\hline
$\mu^+\mu^-$&
$\sqrt{s}$
&$\sigma_{B}$
& $\sigma_{h}$ & $\sigma_{v+s}$ & $\sigma_{v+s+h} $ & $\sigma_{\rm pairs}$
\\
\hline
KLOE     & 1.020 & 542.663(6)  & 1.49406(3)  & -1.7356(2)  & -0.2415(2)
		     &  0.246(7)$\cdot 10^{-7}$ \\
BES    & 3.097 & 173.98(2)   & 1.01652(3)  & -1.09665(1) & -0.08013(3) &  0.001337(5)         \\
BES    & 3.650 & 129.0958(4) & 0.83245(2)  & -0.88149(1) & -0.04904(2) &  0.002003(6)  \\
BES    & 3.686 & 123.32(1)   & 0.82215(2)  & -0.86988(1) & -0.04773(2) &  0.002035(6)  \\
BaBar    & 10.56 & 5.481(1)    & 0.075789(2) & -0.079231(2)& -0.003442(3)&  0.000451(2)      \\
Belle    & 10.58 & 6.73555(4)  & 0.080377(8) & -0.09009(1) & -0.00971(1) &  0.0007587(14) \\
\hline
\end{tabular}
\end{center}
\end{table}

\begin{table}[h]                   
\setlength{\tabcolsep}{0.3pc}
\caption[]{\emph{
Results for tau pair corrections at different energies, in GeV,
for the reference event selection defined in Appendix A.
The $\sigma_{B}$ is the Born cross section within the acceptance cuts.
The separation of soft and hard photons is fixed at
$\omega / E_{\rm beam}=10^{-4}$,
where $\omega = E_{\gamma}^{\min}$ (soft photon cut-off).
All cross sections are given in nb.
}}
\label{table-NNLO-tau}
\begin{center}
\begin{tabular}{lcccccc}
\hline
$\tau^+\tau^-$&
$\sqrt{s}$
&$\sigma_{B}$
& $\sigma_{h}$ & $\sigma_{v+s}$ & $\sigma_{v+s+h} $ & $\sigma_{\rm pairs}$
\\
\hline
KLOE     & 1.020 & 542.663(6)  & 0.0201637(4) & -0.023412(2)  & -0.003248(2)   & 0  \\
BES    & 3.097 & 173.98(2)   & 0.049672(2)  & -0.0540(1)    & -0.0044(1)     & 0  \\
BES    & 3.650 & 129.0958(4) & 0.058674(2)  & -0.0633(1)    & -0.0046(1)     & 0  \\
BES    & 3.686 & 123.32(1)   & 0.057923(2)  & -0.0622(1)    & -0.0043(1)     & 0  \\
BaBar    & 10.56 & 5.481(1)    & 0.0138398(4) & -0.0144654(2) &
		     -0.0006257(5)  & 0.120(3) $\cdot 10^{-8}$ \\
Belle    & 10.58 & 6.73555(4)  & 0.014428(4)  & -0.01602(1)   & -0.00159(1)    & 0.0000321(1)  \\
\hline
\end{tabular}
\end{center}
\end{table}

\begin{table}[h]
\setlength{\tabcolsep}{0.3pc}
\caption[]{\emph{
Results for pion pair corrections at different energies, in GeV,
for the reference event selection defined in Appendix A.
The $\sigma_{B}$ is the Born cross section within the acceptance cuts.
The separation of soft and hard photons is fixed at
$\omega / E_{\rm beam}=10^{-4}$,
where $\omega = E_{\gamma}^{\min}$ (soft photon cut-off).
All cross sections are given in nb.
}}
\label{table-NNLO-pi}
\begin{center}
\begin{tabular}{lcccccc}
\hline
$\pi^+\pi^-$&
$\sqrt{s}$
&$\sigma_{B}$
& $\sigma_{h}$ & $\sigma_{v+s}$ & $\sigma_{v+s+h} $ & $\sigma_{\rm pairs}$
\\
\hline
KLOE     & 1.020 & 542.663(6)  & 1.17402(8)  & -1.35988(2)  & -0.18586(8)   & 0  \\
BES    & 3.097 & 173.98(2)   & 0.95919(3)  & -1.03394(3)     & -0.07475      & 0.000153(2)  \\
BES    & 3.650 & 129.0958(4) & 0.77337(2)  & -0.81806(3)     & -0.04469      & 0.000539(7)  \\
BES    & 3.686 & 123.32(1)   & 0.76286(2)  & -0.80626(3)     & -0.04340      & 0.000564(8)  \\
BaBar    & 10.56 & 5.481(1)    & 0.051037(2) & -0.053328(3) & -0.002291(4)  & 0.000029(3)  \\
Belle    & 10.58 & 6.73555(4)  & 0.054457(6) & -0.0612(1)  & -0.0067(1)   & 0.00015(1)   \\
\hline
\end{tabular}
\end{center}
\end{table}

\begin{table}[h]
\setlength{\tabcolsep}{0.3pc}
\caption[]{\emph{
Results for hadronic corrections at different energies, in GeV,
for the reference event selection defined in Appendix A.
The $\sigma_{B}$ is the Born cross section within the acceptance cuts.
The separation of soft and hard photons is fixed at
$\omega / E_{\rm beam}=10^{-4}$,
where $\omega = E_{\gamma}^{\min}$ (soft photon cut-off).
All cross sections are given in nb.
}}
\label{table-NNLO-had}
\begin{center}
\begin{tabular}{lcccccc}
\hline
hadrons &
$\sqrt{s}$
&$\sigma_{B}$
& $\sigma_{h}$ & $\sigma_{v+s}$ & $\sigma_{v+s+h} $ & $\sigma_{\rm{pion \ pair \ only}}$
\\
\hline
KLOE   & 1.020 &542.663(6)   &  1.5248(6)   &-1.062(8) &0.463(8)  & 0  \\
BES    & 3.650 &129.0958(4)  &  1.66065(8)  &-1.81(1)  &-0.15(1)      & 0.000539(7)  \\
BaBar  & 10.56 &5.481(1)     &  0.17995(2)  &-0.1888(4) &-0.0088(4) & 0.000029(3)  \\
Belle  & 10.58 &6.73555(4)   &  0.18969(1)  &-0.2124(5) &-0.0227(5) & 0.00015(1)  \\
\hline
\end{tabular}
\end{center}
\end{table}

As one can see from Tables \ref{table-el-NNLO}-\ref{table-NNLO-had} the
discussed corrections cannot be ignored for high 
precision luminosity measurements. 
Actually the relative size of electron pair corrections amount to about 
0.3\% at KLOE, $0.1-0.2$\% at BES and BaBar, and below the $10^{-3}$  level
at Belle only. The contribution of muon pair and hadronic corrections is
slightly smaller, in the 0.05-0.1\% range at all meson factories, while
the contribution of tau pair corrections is, not surprisingly, generally
negligible.Therefore, the pair corrections have to be implemented into 
a MC event generator at least in an approximated form (see the
next Section for their implementation in the generator \textsc{BabaYaga@NLO}). 
In particular, for
the real emission one can conclude that only the reaction
$e^+e^- \to e^+e^-e^+e^-$ gives significant contributions to the 
cross section used in the luminosity measurements. When the accuracy of
 the experiment reaches the level $10^{-3}$ this process has to be
 considered and its contributions added to the theoretical cross section
 or alternatively subtracted as a background from the experimental cross
 section. 

\section{The NNLO massive corrections in \textsc{BabaYaga@NLO}  \label{sec-babayaga}}

\subsection{The program}
\label{theprogram}

BabaYaga is an event generator for precise simulations of the processes 
$e^+ e^- \to e^+ e^-, \mu^+\mu^-,\gamma\gamma$ in QED. It was developed for precision measurements with per 
mille accuracy of the luminosity of GeV scale $e^+ e^-$ colliders. It has been adopted and is 
still presently used for this purpose by KLOE, BES/BES-III, CLEO, BaBar and Belle experiments. 

BabaYaga is released in two versions, the more accurate \textsc{BabaYaga@NLO} and BabaYaga 3.5,
{\tt http://www2.pv.infn.it/$\sim$hepcomplex/babayaga.html}. In BabaYaga 3.5 the most relevant QED radiative corrections are 
included in a pure Parton Shower (PS) approach, even though improved to include 
radiation interference effects for a more accurate description of radiative events. \textsc{BabaYaga@NLO}
also includes non-logarithmically enhanced 
$O(\alpha)$ corrections, which were the main source of theoretical error of BabaYaga 3.5. 
As summarised in Section \ref{general}, the necessary NLO ingredients are matched 
in \textsc{BabaYaga@NLO} with the PS approach, in order to preserve exponentiation of large 
contributions and ensure normalisation at NLO accuracy.

Concerning the NNLO corrections that are the concern of the present study, the contribution of 
NNLO massive corrections included in the code 
is approximate and comes, as detailed in Section \ref{pairs-babayaga}, from

\begin{itemize}
\item[$a)$]  insertion of self-energy corrections in NLO virtual + soft photon correction;
\item[$b)$]  insertion of self-energy corrections in NLO hard photon correction.
\end{itemize}

From a theoretical point of view, the part $b)$ coincides with the contribution to the NNLO corrections described in Section \ref{sec-hard}, 
although the $e^+ e^- \to e^+ e^- \gamma$ matrix element, phase space and related MC integration are completely independent 
of the calculation previously discussed. From a numerical point of view, one should expect {\it a priori} that the two calculations 
provide results in agreement within the respective statistical uncertainties, as it will be studied in the following. On the other hand, 
the contribution $a)$ is just a subset of the full NNLO correction described in Section \ref{sec-virt-nnlo}, as 
further detailed in Section \ref{pairs-babayaga}.

Furthermore, the contributions due to real pair emission of the type $2 \to 2 + (2)$, which are included in the complete NNLO benchmark calculation
(Sections \ref{sec-e}-\ref{sec-p}), are presently neglected in the program.

\subsection{General formulation of \textsc{BabaYaga@NLO}}
\label{general}

As far as photon corrections are concerned, the two basic ingredients combined 
in \textsc{BabaYaga@NLO} to guarantee the target theoretical accuracy are
\begin{enumerate}
\item exact NLO QED corrections (soft+virtual and hard contributions);
\item leading QED logarithms due to multiple collinear and soft radiation beyond $O(\alpha)$.
\end{enumerate}
The matching is performed according to the following formula~\cite{Balossini:2006wc}
\begin{equation}
d\sigma^{\infty}_{\rm matched}=
F_{SV}~\Pi(Q^2,\varepsilon)~
\sum_{n=0}^\infty \frac{1}{n!}~
\left( \prod_{i=0}^n F_{H,i}\right)~
|{\cal M}_{n,LL}|^2~
d\Phi_n
\label{matchedinfty}
\end{equation}
where
\begin{enumerate}

\item $\Pi(Q^2,\varepsilon)$ is the Sudakov form factor. It describes
      universal (process independent) virtual + soft radiation up to the
      energy fraction $\epsilon$ (soft-hard separator)
\begin{eqnarray*}
&&\Pi(Q^2,\varepsilon)=
\exp\left(
-\frac{\alpha}{2\pi}~I_+~L^\prime
\right),~~~~L^\prime=\log\frac{Q^2}{m_e^2},~~~~
I_+\equiv
\int_0^{1-\varepsilon}
dz P(z) \nonumber\\
&&P(z) = \frac{1+z^2}{1-z}.
\end{eqnarray*}
In the program, the scale $Q^2$ is chosen such that $L^\prime=\log\frac{Q^2}{m_e^2} = \log\frac{st}{u m_e^2} - 1\equiv L-1$ where $s$, $t$ and $u$ are the Mandelstam variables of the process and $m_e$ is the electron mass. This choice is dictated by the perturbative NLO calculation of the Bhabha process and ensures that, in addition to initial and final state leading contributions, also initial-final state interference leading effects are resummed to all orders.

\item $|{\cal M}_{n,LL}|^2$: $n$-photon emission squared amplitude in collinear approximation, with phase space factor $d\Phi_n$

\item $F_{SV}$ and $F_H$: residuals of the exact NLO calculation w.r.t. the leading log approximation ensured by the ingredients 1. and 2. above, i.e. 
\begin{eqnarray*}
&&F_{SV} = 1+\left(C_\alpha-C_{\alpha,LL}\right), ~~~~F_H~=~1+\frac{|{\cal M}_1|^2-|{\cal M}_{1,LL}|^2}{|{\cal M}_{1,LL}|^2} \nonumber\\
&&C_{\alpha}: {\rm exact} \, {\rm soft \, plus \, virtual \, NLO} \, K \, {\rm factor},  \nonumber\\
&&C_{\alpha,LL}:  {O}(\alpha) \, {\rm \, expansion \, of} \, {\rm the \, Sudakov \, form \, factor}  \nonumber\\
&&|{\cal M}_1|^2: {\rm exact \, NLO \, hard \, bremsstrahlung \, squared \, matrix \, element}
\label{FSVH}
\end{eqnarray*}

\end{enumerate}
The explicit expressions of all the above contributions can be found in \cite{Balossini:2006wc}.

A further necessary ingredient is the correction due to vacuum polarisation. It is included in the Bhabha Born matrix 
element setting $r_s=\alpha(s)/\alpha$ and $r_t=\alpha(t)/\alpha$ and rescaling the $s$ and $t$ channel amplitudes as follows
\begin{equation}
|{\cal M}_0|^2=|{\cal M}_{0,s} + {\cal M}_{0,t}|^2~~\to~~
|{\cal M}_{0,VP}|^2 = |{\cal M}_{0,s}r_s + {\cal M}_{0,t}r_t|^2 \, .
\label{vpborn}
\end{equation}
In the code, we use the resummed expression $\alpha(q^2)=\alpha/(1-\Delta\alpha(q^2))$, where $\Delta\alpha(q^2)$
is the fermionic contribution to the photon self-energy. It is treated analytically for the leptonic
and top-quark one-loop contributions, while the non-perturbative five quark (hadronic)
contribution, $\Delta\alpha^{(5)}_{had}$, is included according to
the latest Jegerlehner \cite{jegerlehner:2010} and Teubner et al.~\cite{teubner:2010} parameterisations.
In order to include an important class of $O(\alpha^2)$ factorizable corrections, we insert the vacuum polarisation correction 
in the NLO cross section too, both in the soft plus virtual contribution and the hard photon matrix element. The inclusion of the 
vacuum polarisation both in the soft plus virtual and hard photon part guarantees that their sum is independent of the 
soft-hard separator $\epsilon$, as we explicitly checked.

\subsection{NNLO massive corrections  in \textsc{BabaYaga@NLO}}
\label{pairs-babayaga}

As detailed in \cite{Balossini:2006wc}, it is possible to extract from the matched formula given in Eq.~(\ref{matchedinfty}) 
the different pieces contributing to the cross section at NNLO. In particular, 
to explain how NNLO massive corrections are (approximately) taken into account in \textsc{BabaYaga@NLO}, let us write the
expansion up to $O(\alpha^2)$ of the cross section with soft plus virtual corrections. It can be derived from the first 
($n = 0$) term of the infinite sum in Eq. (\ref{matchedinfty}). In order to highlight the $s$,
$t$ and interference contributions, first we define
\begin{eqnarray}
d\sigma_{0} &=& d\sigma_{s,0}
+d\sigma_{t,0}+d\sigma_{st,0}
\nonumber \\
&\equiv& (B_s+B_t+B_{st})d\sigma_0 ,
\nonumber\\
d\sigma^\alpha_{SV} &=& d\sigma^\alpha_{s,SV}
+d\sigma^\alpha_{t,SV}+d\sigma^\alpha_{st,SV}
\nonumber \\
&\equiv&(E_s+E_t+E_{st})d\sigma_0 ,
\label{oa2}
\end{eqnarray}
where $B_s$, $B_t$ and $B_{st}$ are the percentage $s$, $t$ and $st$ 
lowest-order contributions to the complete Bhabha differential cross section $d\sigma_0$, 
and $E_s$, $E_t$ and $E_{st}$ are the NLO SV correction factors for each contribution, in units of 
 $d\sigma_0$. Truncating every factor in Eq.~(\ref{oa2}), improved with vacuum
polarisation effects as described above,  we get  from Eq.~(\ref{matchedinfty}) at $O(\alpha^2)$
\begin{eqnarray}
\frac{d\sigma_{SV}}{d\sigma_0}
   &\simeq& \left(1 + V + \frac{V^2}{2}\right) 
\nonumber \\
   && \times~\left[ 1+ (E_s - V B_s) r_s^2 + (E_t - V B_t) r_t^2 
                 +(E_{st} - V B_{st}) r_s r_t \right] 
\nonumber \\
   && \times~\left( B_s r_s^2 + B_t r_t^2 + B_{st} r_s r_t \right) ,
\label{beforeexpansion}
\end{eqnarray}
where $V = -(2 \alpha /\pi) I_+ L'$ is the $O(\alpha)$ truncation of
the Sudakov form factor, 
$E_i$ and $B_i$ have been defined above and
$r_{s,t}$ are the vacuum 
polarisation corrections for the $s$ and $t$ 
channels, respectively. If we define $1/(1 -
\Delta\alpha(q^2))\equiv1/(1-\delta_{q^2})$, the $r^2_S$, $r^2_t$ and
$r_sr_t$ read
\begin{eqnarray}
r_s^2 &=& 1 + 2 \delta_s + 3 \delta_s^2 , \nonumber \\
r_t^2 &=& 1 + 2 \delta_t + 3 \delta_t^2 , \nonumber \\
r_sr_t &=& 1 + \delta_s +\delta_t +\delta_s^2 +\delta_t^2 +\delta_s\delta_t  .
\end{eqnarray}
Retaining only terms up to $O(\alpha^2)$, Eq.~(\ref{beforeexpansion}) 
reads 
\begin{eqnarray}
\frac{d\sigma_{SV}}{d\sigma_0}
&=& 
1 \nonumber \\
&&+~ V  + (E_s - V B_s) + (E_t - V B_t) + (E_{st} - V B_{st}) \nonumber \\
&&+~ 2 (B_s \delta_s + B_t \delta_t) + B_{st} (\delta_s + \delta_t) \nonumber \\
&&+~ 1/2{V^2} \nonumber \\
&&+~  (E_s - V B_s) \delta_s + (E_t - V B_t) \delta_t 
+ (E_{st} - V B_{st}) (\delta_s + \delta_t )\nonumber \\
&&+~ 3 (B_s \delta_s^2 + B_t \delta_t^2) + B_{st} 
(\delta_s^2 + \delta_t^2 + \delta_s \delta_t) \nonumber \\
&&+~ V [(E_s - V B_s) + (E_t - V B_t) + (E_{st} - V B_{st}] \nonumber \\
&&+~ V [2 (B_s\delta_s + B_t\delta_t) + B_{st}(\delta_s +\delta_t)] \nonumber \\
&&+~ \left[(E_s-V B_s)+(E_t-V B_t)+(E_{st} - V B_{st})\right]
\nonumber \\
&&\times~\left[2(B_s\delta_s+B_t\delta_t)+B_{st} (\delta_s + \delta_t)\right] .
\label{expansion}
\end{eqnarray}
The first line of Eq.~(\ref{expansion}) is the Born contribution, the
second line is the one soft photon plus  one loop virtual correction
(notice that it is equal to $E_s+E_t+E_{st}$ because
$B_s+B_t+B_{st}=1$), the third line is the vacuum polarisation
correction at $O(\alpha)$ and the remaining lines represent the cross section with 
$O(\alpha^2)$ soft plus virtual corrections. From the latter it is simple to disentangle the NNLO contribution due to the insertion of the vacuum polarisation 
correction in the $O(\alpha)$ soft plus virtual coefficients. The relevant terms are those containing a $\delta_i$ factor. Among 
them, there is a pure two-loop self-energy correction (sixth line in Eq.~(\ref{expansion})), which we discard for the comparison 
with the exact NNLO calculation, in accordance with the discussion of Section \ref{sec-virt-nnlo}. Therefore the formula of interest reduces to
\begin{eqnarray}
\frac{d\sigma_{SV}}{d\sigma_0}
&=& 
  (E_s - V B_s) \delta_s + (E_t - V B_t) \delta_t 
+ (E_{st} - V B_{st}) (\delta_s + \delta_t )\nonumber \\
&&+~ V [2 (B_s\delta_s + B_t\delta_t) + B_{st}(\delta_s +\delta_t)] \nonumber \\
&&+~ \left[(E_s-V B_s)+(E_t-V B_t)+(E_{st} - V B_{st})\right]
 \nonumber \\
&&\times~\left[2(B_s\delta_s+B_t\delta_t)+B_{st} (\delta_s + \delta_t)\right] .
\label{expansionpair}
\end{eqnarray}
Equation~(\ref{expansionpair}) is used in the present study to validate the approximate treatment of NNLO massive 
corrections as in \textsc{BabaYaga@NLO} in the soft plus virtual regime.

Equation (\ref{expansionpair}) must be added to the contribution obtained by dressing the hard bremsstrahlung cross section 
with self-energy corrections. The hard photon matrix element is the sum of eight amplitudes where the real photon is
attached to a $s$ or $t$ channel-like diagram. As in Eq.~(\ref{vpborn}), those amplitudes are rescaled by $r_s$ and $r_t$,
respectively, to account for the effect of vacuum polarisation. In
\textsc{BabaYaga@NLO}, the squared amplitude for the emission of a real photon
$|{\cal M}_1|^2$ is exact as calculated through $\tt FORM$
\cite{Vermaseren:2000nd} and cross checked with the output of
the $\tt ALPHA$ algorithm~\cite{Caravaglios:1995cd}.

To summarize, the best knowledge of running $\alpha_{\rm{\ssQ\ssE\ssD}}$ and all the virtual factorizable NNLO vacuum polarization corrections are included 
in \textsc{BabaYaga@NLO}. In particular, within the full set of (reducible and irreducible) NNLO virtual massive corrections present in the exact
calculation described in Section \ref{sec-virt-nnlo}, only the subset of loop-by-loop corrections (see Figure 2.2) is taken into account in the code. 
In other words, the contributions of purely irreducible NNLO vertex and box corrections is not included. However, since the real pair 
emission corrections are neglected as well, this means that there is no imbalance of $\ln^3(s/m_e)$ terms in \textsc{BabaYaga@NLO}, 
thus respecting the compensation mechanism of leading mass singularities discussed in Section \ref{sec-virt-nnlo} and Section \ref{sec-e} 
(see Eq.~(\ref{sig-irr-vert}) and Eq.~(\ref{sp})).

\subsection{Numerical results of \textsc{BabaYaga@NLO} \label{sec-baba-3}}

Below we give the benchmark results from the \textsc{BabaYaga@NLO} MC
event generator. Their detailed comparison to the exact results will be
given in the next Section.

\begin{table}[h]
\setlength{\tabcolsep}{0.3pc}
\caption[]{\emph{
\textsc{BabaYaga@NLO} results for electron pair corrections at different energies, in GeV,
for the reference event selection defined in Appendix A.
The $\sigma_{B}$ is the Born cross section within the acceptance cuts.
The separation of soft and hard photons is fixed at
$\omega / E_{\rm beam}=10^{-4}$,
where $\omega = E_{\gamma}^{\min}$ (soft photon cut-off).
All cross sections are given in nb.
}
}
\label{table-el-babayaga}
\begin{center}
\begin{tabular}{lcccccc}
\hline
$e^+e^-$& $\sqrt{s}$ &$\sigma_{B}$ & $\sigma_{h}$ & $\sigma_{v+s}$ & $\sigma_{v+s+h} $ & $\sigma_{\rm pairs}$
\\
\hline
KLOE   & 1.020 & 542.663(6) &9.5022(8)  &-11.0721(4)   &-1.5699(9)    &-   \\
BES    & 3.097 & 173.98(2)  &4.1624(4)  &-4.4818(2)    &-0.3194(5)    &-   \\
BES    & 3.650 & 129.0958(4)&3.1960(3)  &-3.3730(2)    &-0.1770(4)    &-   \\
BES    & 3.686 & 123.32(1)  &3.1447(3)  &-3.3163(2)    &-0.1716(4)    &-   \\
BaBar  & 10.56 & 5.481(1)   &0.20244(2) &-0.20971(5)   &-0.00727(5)   &-   \\
Belle  & 10.58 & 6.73555(4) &0.21563(2) &-0.23994(2)   &-0.02431(3)   &-   \\
\hline
\end{tabular}
\end{center}
\end{table}

\begin{table}[h]
\setlength{\tabcolsep}{0.3pc}
\caption[]{\emph{
\textsc{BabaYaga@NLO} results for muon pair corrections at different energies, in GeV,
for the reference event selection defined in Appendix A.
The $\sigma_{B}$ is the Born cross section within the acceptance cuts.
The separation of soft and hard photons is fixed at
$\omega / E_{\rm beam}=10^{-4}$,
where $\omega = E_{\gamma}^{\min}$ (soft photon cut-off).
All cross sections are given in nb.
}
}
\label{table-mu-babayaga}
\begin{center}
\begin{tabular}{lcccccc}
\hline
$\mu^+\mu^-$&
$\sqrt{s}$
&$\sigma_{B}$
& $\sigma_{h}$ & $\sigma_{v+s}$ & $\sigma_{v+s+h} $ & $\sigma_{\rm pairs}$
\\
\hline
KLOE   & 1.020 & 542.663(6) &1.4942(2)   &-1.7441(2)    &-0.2499(3)  &-   \\
BES    & 3.097 & 173.98(2)  &1.01672(9)  &-1.0960(2)    &-0.0793(2) &-   \\
BES    & 3.650 & 129.0958(4)&0.83252(7)  &-0.88041(9)   &-0.0479(1)  &-   \\
BES    & 3.686 & 123.32(1)  &0.82221(7)  &-0.8688(1)    &-0.0466(1)  &-   \\
BaBar  & 10.56 & 5.481(1)   &0.07580(1)  &-0.07872(2)   &-0.00292(2) &-   \\
Belle  & 10.58 & 6.73555(4) &0.080376(6) &-0.08948(2)   &-0.00910(2) &-   \\
\hline
\end{tabular}
\end{center}
\end{table}

\begin{table}[h]
\setlength{\tabcolsep}{0.3pc}
\caption[]{\emph{
\textsc{BabaYaga@NLO} results for tau pair corrections at different energies, in GeV,
for the reference event selection defined in Appendix A.
The $\sigma_{B}$ is the Born cross section within the acceptance cuts.
The separation of soft and hard photons is fixed at
$\omega / E_{\rm beam}=10^{-4}$,
where $\omega = E_{\gamma}^{\min}$ (soft photon cut-off).
All cross sections are given in nb.
}
}
\label{table-tau-babayaga}
\begin{center}
\begin{tabular}{lcccccc}
\hline
$\tau^+\tau^-$&
$\sqrt{s}$
&$\sigma_{B}$
& $\sigma_{h}$ & $\sigma_{v+s}$ & $\sigma_{v+s+h} $ & $\sigma_{\rm pairs}$
\\
\hline
KLOE   & 1.020 & 542.663(6) &0.020166(3) &-0.023704(2)  &-0.003538(4)  &-   \\
BES    & 3.097 & 173.98(2)  &0.049683(5) &-0.05421(1)   &-0.00453(1)   &-  \\
BES    & 3.650 & 129.0958(4)&0.058679(7) &-0.06323(2)   &-0.00455(2)   &-   \\
BES    & 3.686 & 123.32(1)  &0.057928(7) &-0.06219(2)   &-0.00426(2)   &-   \\
BaBar  & 10.56 & 5.481(1)   &0.013847(4) &-0.014541(4)  &-0.000694(6)  &-   \\ 
Belle  & 10.58 & 6.73555(4) &0.014423(1) &-0.016091(7)  &-0.001668(7)  &-   \\
\hline
\end{tabular}
\end{center}
\end{table}

\begin{table}[h]
\setlength{\tabcolsep}{0.3pc}
\caption[]{\emph{
\textsc{BabaYaga@NLO} results for hadronic corrections at different energies, in GeV,
for the reference event selection defined in Appendix A.
The $\sigma_{B}$ is the Born cross section within the acceptance cuts.
The separation of soft and hard photons is fixed at
$\omega / E_{\rm beam}=10^{-4}$,
where $\omega = E_{\gamma}^{\min}$ (soft photon cut-off).
All cross sections are given in nb.
}
}
\label{table-pi-babayaga}
\begin{center}
\begin{tabular}{lcccccc}
\hline
hadrons&
$\sqrt{s}$
&$\sigma_{B}$
& $\sigma_{h}$ & $\sigma_{v+s}$ & $\sigma_{v+s+h} $ & $\sigma_{\rm hadrons}$
\\
\hline
KLOE   & 1.020 &542.663(6)   &1.5247(5)  &-1.126(2)    &0.399(2)    &-   \\
BES    & 3.650 &129.0958(4)  &1.6613(3)  &-1.7860(2)   &-0.1247(4)  &-   \\
BaBar  & 10.56 &5.481(1)     &0.17984(2) &-0.18760(4)  &-0.00776(5) &-   \\
Belle  & 10.58 &6.73555(4)   &0.18964(3) &-0.21089(5)  &-0.02125(6) &-   \\
\hline
\end{tabular}
\end{center}
\end{table}

We repeat here that \textsc{BabaYaga@NLO} does not generate the real pair emission contribution. 
This explains why the corresponding predictions for the cross section $\sigma_{\rm pairs}$
and $\sigma_{\rm hadrons}$ are not given in Tables \ref{table-el-babayaga}-\ref{table-pi-babayaga}.
However, 
 from Section \ref{sec-numNNLO} (see the next Section for a more detailed
 discussion)
 it is clear that only the reaction 
 $e^+e^- \to e^+e^- e^+e^- $ gives a contribution which is relevant
  and needs a particularly careful treatment.
 To account for this reaction a dedicated generator should be used 
 (for example \textsc{Helac--Phegas}).  
 
 Comparing the exact NNLO results of Tables \ref{table-el-NNLO}-\ref{table-NNLO-had}
 with those of \textsc{BabaYaga@NLO} one can note that, while differences 
 are present as expected, in the soft + virtual cross section, the independent 
 predictions of the two calculations for the hard photonic correction 
 $\sigma_{h}$ are in agreement within the MC statistical uncertainty.

\section{\textsc{BabaYaga@NLO} versus the exact NNLO massive corrections\label{sec-num}}
\begin{table}[h]
\setlength{\tabcolsep}{0.3pc}
\caption[]{\emph{
Comparison of the exact massive NNLO with \textsc{BabaYaga@NLO} results 
given for the reference event selections at different meson factories.
The event selections are defined in the Appendix A.
The $\sigma_{\rm{\ssB\ssY}}$ is the  cross section in nb from \textsc{BabaYaga@NLO}, and
 $S_{x}=\frac{\sigma_{x}^{\rm{\ssN\ssN\ssL\ssO}}}{\sigma_{\rm{\ssB\ssY}}}$ with $x=e^+e^-,  lep,
 tot$, where $tot$ stands for leptonic ($ lep$) + hadronic corrections.
}}
\label{table-compar-NNLO-net}
\begin{center}
\begin{tabular}{lclccccc}
\hline
 & $\sqrt{s}$ &  & $\sigma_{\rm{\ssB\ssY}}$ & $S_{e^+e^-}$ [\textperthousand] &
 $S_{lep}$
[\textperthousand] &$S_{had}$ [\textperthousand]&
 $S_{tot}\textperthousand$ [\textperthousand]
\\
\hline
KLOE   & 1.020 & NNLO    &     &-3.935(4) &-4.472(4)  &1.02(2)    &-3.45(2)    \\
       &       &\textsc{BabaYaga@NLO}&   455.71        &-3.445(2) &-4.001(2)  &0.876(5)   &-3.126(5)   \\
BES    & 3.097 & NNLO    &     &-2.246(8) &-2.771(8)  &-          &-           \\
       &       &\textsc{BabaYaga@NLO} &  158.23         &-2.019(3) &-2.548(3)  &-          &-           \\
BES    & 3.650 & NNLO    &     &-1.469(9) &-1.913(9)  &-1.3(1)    &-3.2(1)     \\
       &       &\textsc{BabaYaga@NLO}&  116.41          &-1.521(4) &-1.971(4)  &-1.071(4)  &-3.042(5)   \\
BES    & 3.686 &  NNLO   &     &-1.435(8) &-1.873(8)  &-          &-           \\
       &       &\textsc{BabaYaga@NLO}&  114.27          &-1.502(4) &-1.947(4)  &-          &-           \\
BaBar  & 10.56 &  NNLO   &      &-1.48(2)  &-2.17(2)   &-1.69(8)   &-3.86(8)    \\
       &       &\textsc{BabaYaga@NLO}&   5.195         &-1.40(1)  &-2.09(1)   &-1.49(1)   &-3.58(2)    \\
Belle  & 10.58 &  NNLO   &      &-4.93(2)  &-6.84(2)   &-4.1(1)    &-10.9(1)    \\
       &       &\textsc{BabaYaga@NLO}&  5.501          &-4.42(1)  &-6.38(1)   &-3.86(1)
			 &-10.24(2)   \\
\hline
\end{tabular}
\end{center}
\end{table}

\begin{figure}[t]
\begin{center}
\hspace*{-0.3cm}
\includegraphics[width=0.95\textwidth]{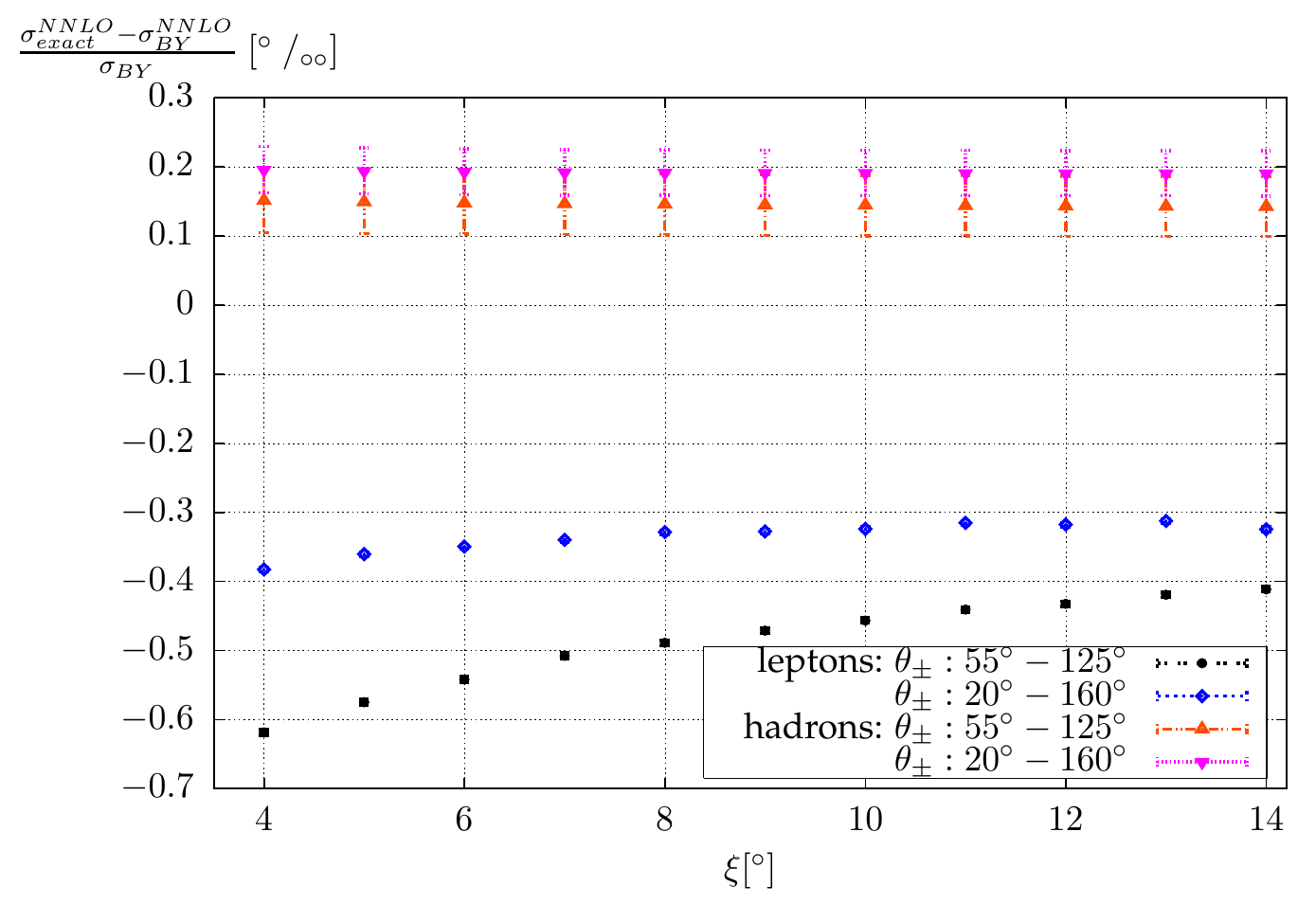}
\caption{\emph{
The relative difference $\frac{\sigma_{\rm exact}^{\rm{\ssN\ssN\ssL\ssO}}-\sigma_{\rm{\ssB\ssY}}^{\rm{\ssN\ssN\ssL\ssO}}}{\sigma_{\rm{\ssB\ssY}}}$ of 
NNLO massive leptonic and hadronic corrections
between exact and \textsc{BabaYaga@NLO},
 as a function of acollinearity cut for two different angular 
 acceptance regions for KLOE-like event selections (see Appendix A).
}
\label{fig-kloe}}
\vspace*{3.0cm}
\end{center}
\end{figure}

\begin{figure}
\begin{center}
\hspace*{-0.3cm}
\includegraphics[width=.95\textwidth]{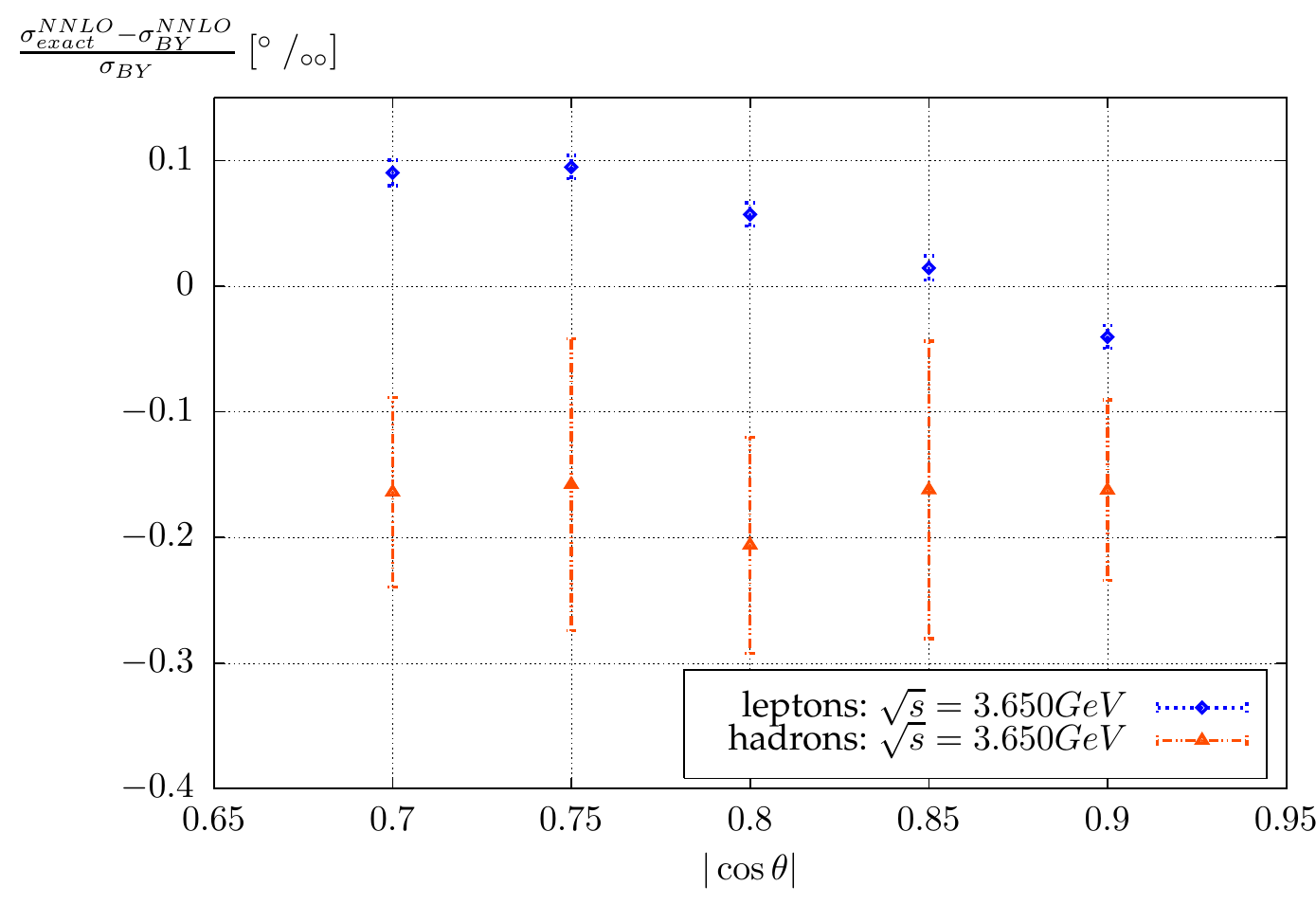}
\caption{\emph{
The relative difference $\frac{\sigma_{\rm exact}^{\rm{\ssN\ssN\ssL\ssO}}-\sigma_{\rm{\ssB\ssY}}^{\rm{\ssN\ssN\ssL\ssO}}}{\sigma_{\rm{\ssB\ssY}}}$ of 
NNLO massive leptonic and hadronic corrections
between exact and \textsc{BabaYaga@NLO}, 
 as a function of an angular cut  for BES-like event selections (see Appendix A).
\label{fig-bes}}}
\end{center}
\end{figure}

\begin{figure}[b]
\begin{center}
\hspace*{-0.1cm}
\includegraphics[width=0.7\textwidth]{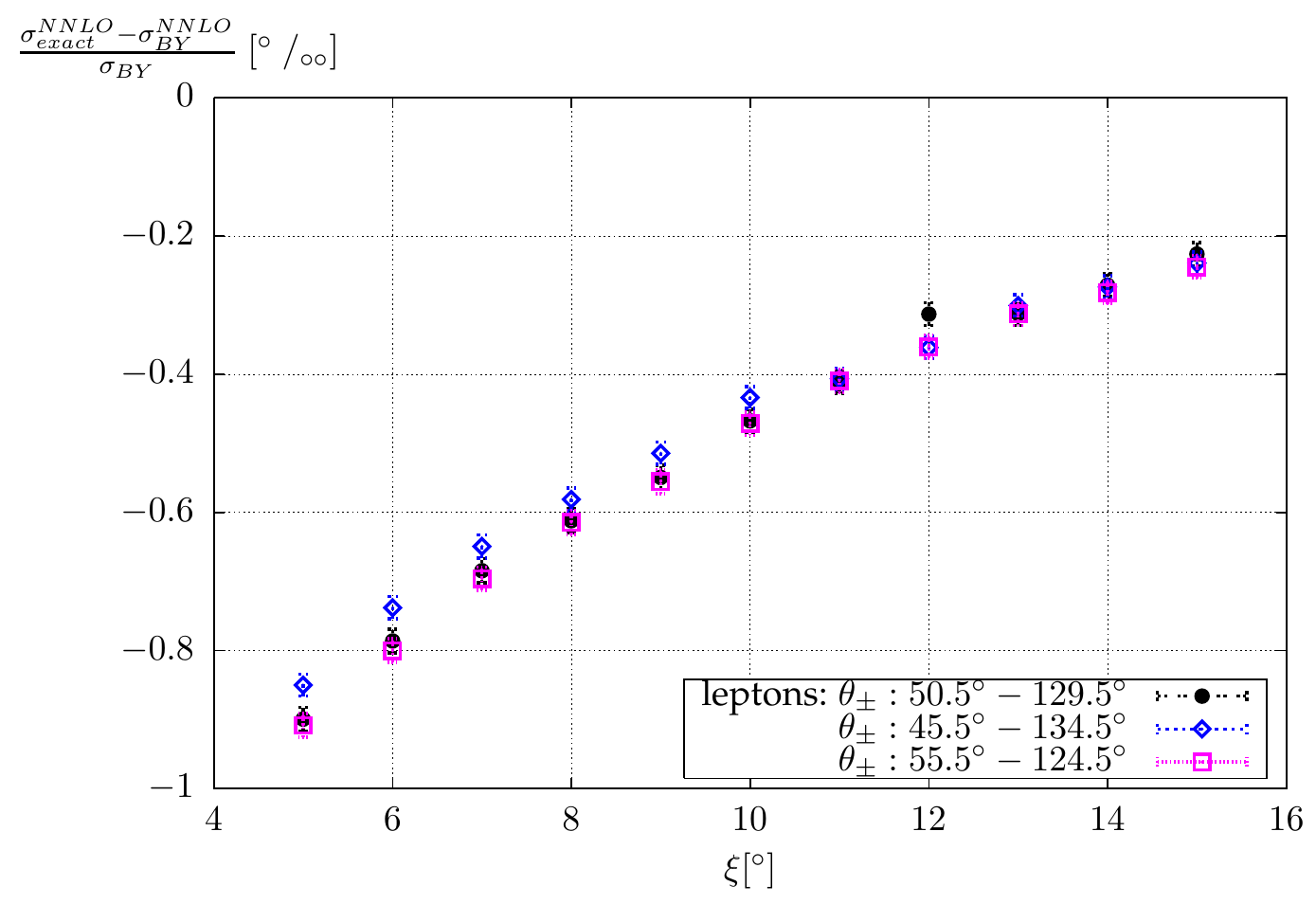}
\includegraphics[width=0.7\textwidth]{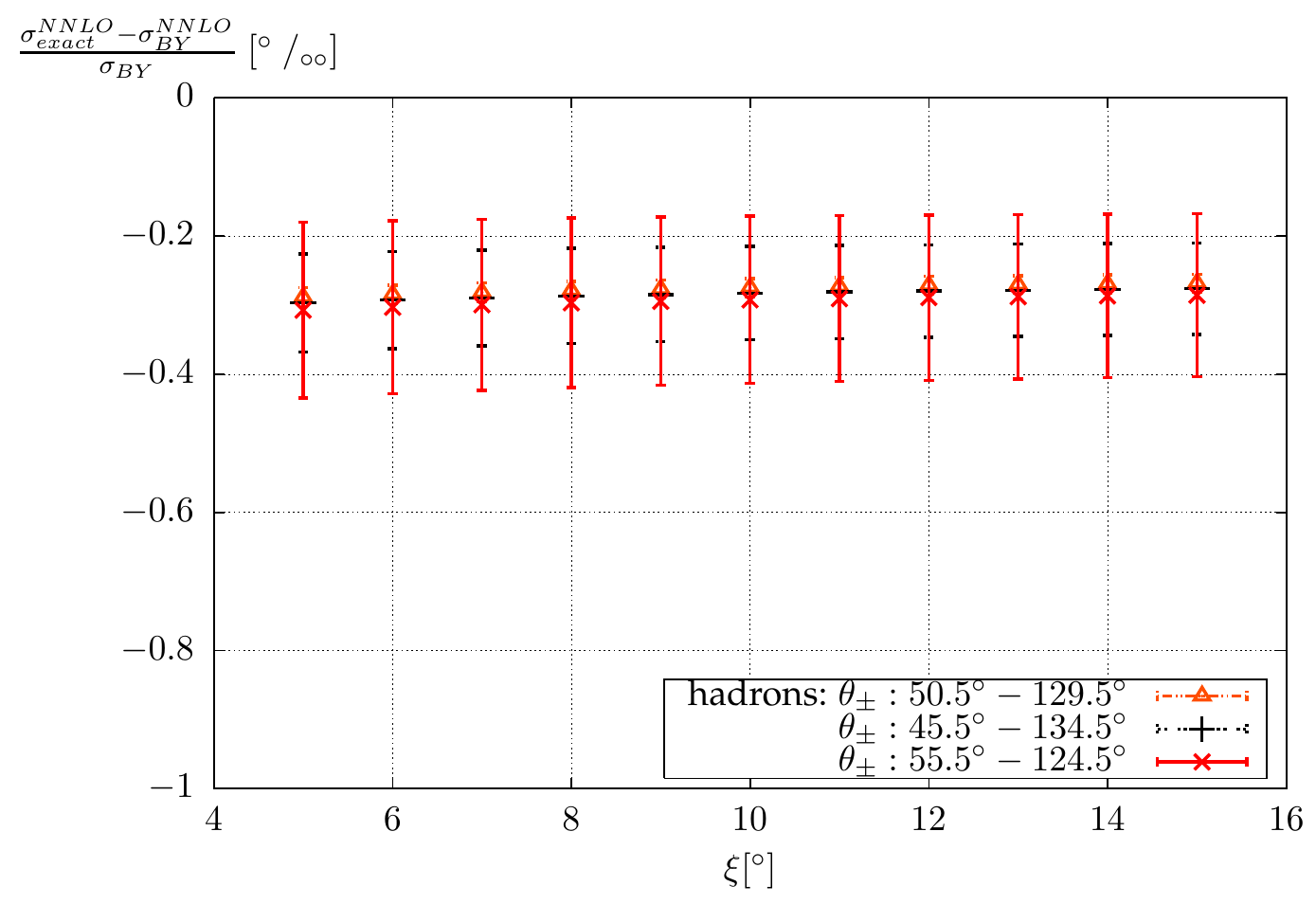}
\caption{\emph{The relative difference $\frac{\sigma_{\rm exact}^{\rm{\ssN\ssN\ssL\ssO}}-\sigma_{\rm{\ssB\ssY}}^{\rm{\ssN\ssN\ssL\ssO}}}{\sigma_{\rm{\ssB\ssY}}}$
of
NNLO massive leptonic (upper plot) and hadronic (lower plot) corrections
between exact and \textsc{BabaYaga@NLO}, 
 as a function of  acollinearity cut for three different angular 
 acceptance regions for Belle-like event selections (see Appendix A).
}
\label{fig-belle}}
\end{center}
\end{figure}

\begin{figure}[tbh]
\begin{center}
\hspace*{-0.1cm}
\includegraphics[width=0.7\textwidth]{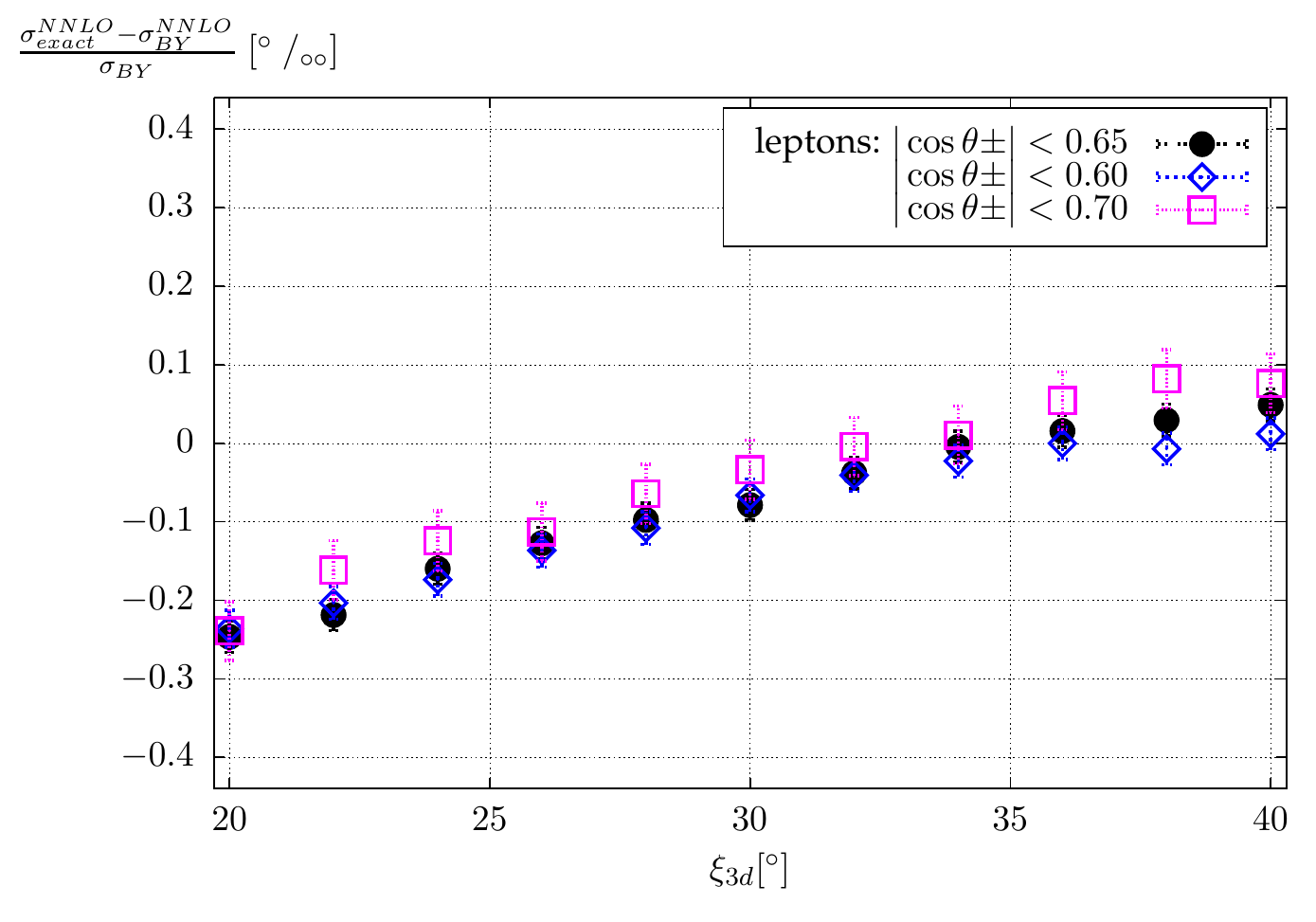}
\includegraphics[width=0.7\textwidth]{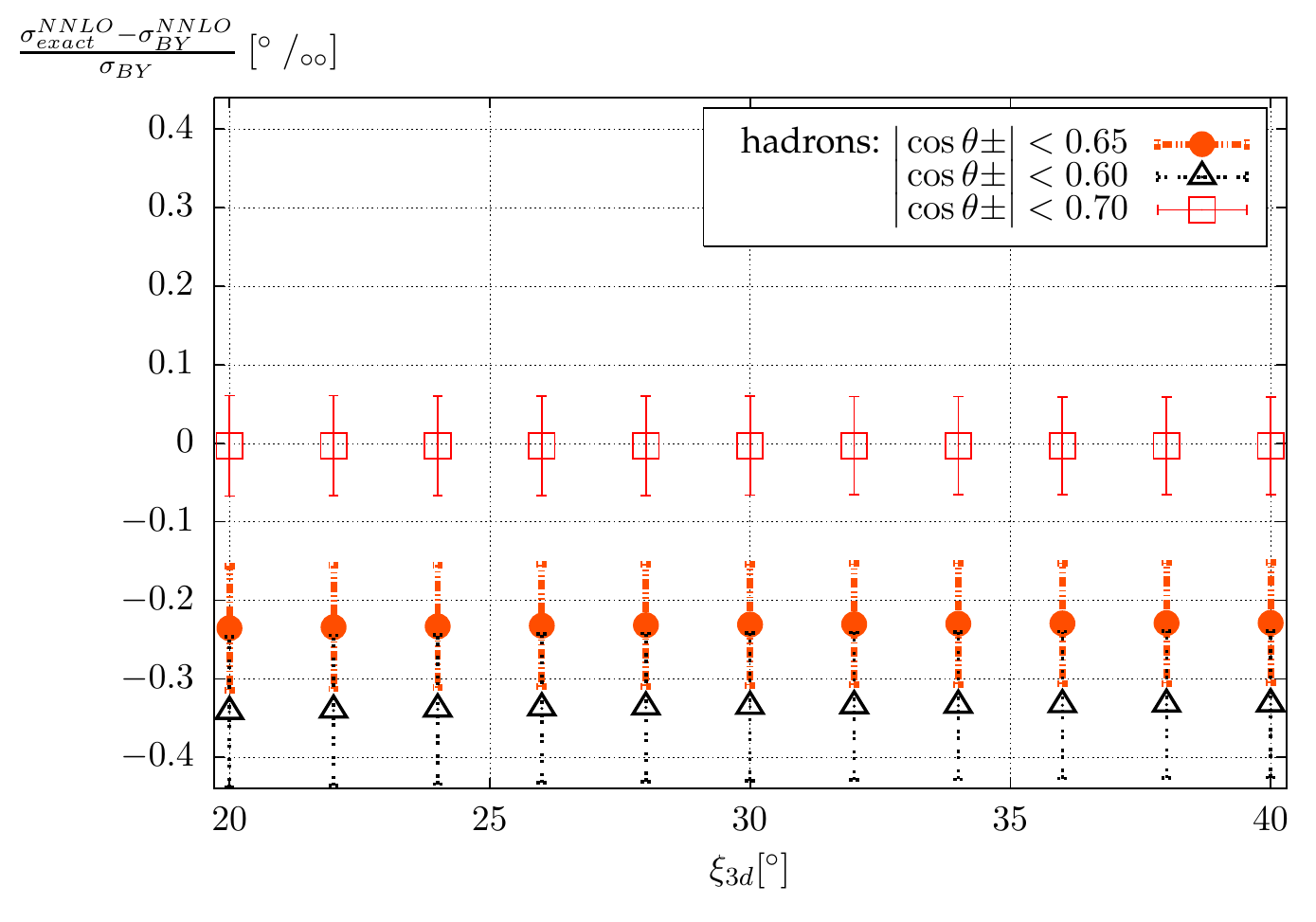}
\caption{\emph{The relative difference 
$\frac{\sigma_{\rm exact}^{\rm{\ssN\ssN\ssL\ssO}}-\sigma_{\rm{\ssB\ssY}}^{\rm{\ssN\ssN\ssL\ssO}}}{\sigma_{\rm{\ssB\ssY}}}$ of NNLO massive 
leptonic (upper plot) and hadronic (lower plot) corrections
between exact and \textsc{BabaYaga@NLO}, 
 as a function of  acollinearity cut for three different angular 
 acceptance regions for BaBar-like event selections (see Appendix A).
\label{fig-babar}}}
\end{center}
\end{figure}

In this Section we would like to answer how well the NNLO massive
corrections are accounted for in the \textsc{BabaYaga@NLO} event generator. The results 
are summarised in Table \ref{table-compar-NNLO-net} and Figs.~\ref{fig-kloe}-\ref{fig-babar}.

In 
Table \ref{table-compar-NNLO-net}
we show the impact of the exact corrections and the corrections given 
 by \textsc{BabaYaga@NLO} for the reference event selections at meson factories.
 One can observe that the leptonic corrections are dominated by
 the electron corrections. Moreover they are well accounted for 
  in the \textsc{BabaYaga@NLO} code with the biggest unaccounted correction of the
 order of 0.5 \textperthousand. In Figs.~\ref{fig-kloe}-\ref{fig-babar}
 we have studied the stability of the obtained results against 
 changes of the event selection. We plot there the relative difference
 between the exact massive NNLO corrections and the \textsc{BabaYaga@NLO} results, 
 i.e. $\frac{\sigma_{\rm exact}^{\rm{\ssN\ssN\ssL\ssO}}-\sigma_{\rm{\ssB\ssY}}^{\rm{\ssN\ssN\ssL\ssO}}}{\sigma_{\rm{\ssB\ssY}}}$, 
 where $\sigma_{\rm{\ssB\ssY}}$ is the full (with all radiative corrections included) 
 prediction of \textsc{BabaYaga@NLO} according to Eq.~(\ref{matchedinfty}).
One can observe that even if for very stringent cuts one can have 
    the differences up to 0.9 \textperthousand~(Belle,
    Fig.\ref{fig-belle}), generally the results are not varying rapidly.
  Comparisons between results for Belle (Fig.\ref{fig-belle}) and BaBar 
  event selections (Fig.\ref{fig-babar}) show also that the actual
  difference is sensitive to the event selection used and it is
  recommended to study the effect on the  NNLO massive corrections,
  whenever the event selection is changed by a given experiment.
 Similar effects are observed for the hadronic corrections (with the
 exception of $J/\psi$ and $\psi(2S)$ energies, which will be examined in a
 separate paper). The biggest hadronic correction missing in \textsc{BabaYaga@NLO}
 is about 0.4 per mille and in addition for KLOE and BES energies the
 missing hadronic contribution is of the opposite sign of the missing
 leptonic contribution, thus partly cancelling each other. For
 $B$-factories the missing leptonic and hadronic corrections are of the 
 same sign, but even there the sum of the missing parts does not exceed
 one per mille (0.7 per mille for the reference event selections).

\section{Conclusions\label{sec-concl}}

  We presented an exact calculation of NNLO massive corrections to
  Bhabha scattering, beyond approximations previously addressed
  in the literature, and performed detailed studies of the missing 
  part of these corrections in the \textsc{BabaYaga@NLO}
  MC program. The
  approximate \textsc{BabaYaga@NLO} formulae were confronted with
   the exact NNLO results for event selections close to the experimental
  ones. The stability of the results with changing of the event
  selections was also examined. 
  
  We found that NNLO massive corrections are relevant for precision luminosity 
  measurements with $10^{-3}$ accuracy, their relative total contribution to the 
  Bhabha scattering cross section being of a few per mille. 
  Hierarchically, the electron pair contribution is the largely dominant part
  of the correction, the muon pair and hadronic correction, which are the next-to-important
  effect, are quantitatively on the same grounds, while the tau pair 
  contribution is negligible for the energies of meson factories.
  
  Thanks to the exact NNLO calculation we tested the theoretical accuracy
  of the generator \textsc{BabaYaga@NLO} which includes such corrections
   according to an approximate treatment through insertion of 
   self-energy contributions into NLO correction factors. On the
   grounds of several numerical results, we concluded that the very bulk
   of the correction is taken into account in the program. For reference
   realistic event selections the maximum observed difference is at the
   level of 0.07\%. When cuts are varied the sum of the missing pieces
   can reach 0.1\%, but for very tight acollinearity cuts only.
   
   As a possible perspective, it is worth mentioning that the leading logarithmic
   part of the missing pieces in \textsc{BabaYaga@NLO}, coming from the interplay
   between NNLO virtual and real pair corrections, could be accounted for 
   by means of appropriate singlet/non-singlet QED structure functions, 
   as discussed e.g. in \cite{Skrzypek:1992vk,Arbuzov:1999cq,Arbuzov:2010zzb} and done 
   in the past for small-angle Bhabha
   scattering at LEP \cite{Jadach:1992nk}. However, this improvement
   of the theoretical formulation of \textsc{BabaYaga@NLO} would be necessary only 
   whenever the experimental precision of the luminosity measurements 
   should require it and should be accompanied by the inclusion of
   other sources of NNLO ingredients presently neglected in the code 
   and contributing at an accuracy below 
   the $10^{-3}$ level \cite{Actis:2010gg,Montagna:2010jm}.
   As  already discussed in the paper,  open issues of the present work
   are a more detailed study of hadronic NNLO corrections and of the uncertainty induced by hadronic vacuum polarisation insertions 
   to Bhabha scattering 
   in a close vicinity of narrow resonances. We plan to address the above
   perspectives in future works.
  
\acknowledgments{
  We would like to thank Thomas Teubner for many useful discussions. 
  We are grateful to Wang Ping for having provided us the details about the cuts used 
  for luminosity monitoring at BEPC II and KEK.
 This work is a part of the activity of the ``Working Group on Radiative
 Corrections and Monte Carlo Generators for Low Energies" 
 [\href{http://www.lnf.infn.it/wg/sighad/}{http://www.lnf.infn.it/wg/sighad/}].
M. Worek acknowledges support by the Initiative and Networking Fund of the German Helmholtz Association "Physics at the Terascale", contract HA-101.
This work is also supported in part by Sonderforschungsbereich/Trans\-re\-gio SFB/TRR 9 of DFG
``Com\-pu\-ter\-ge\-st\"utz\-te Theoretische Teil\-chen\-phy\-sik", 
European Initial Training Network LHCPHENOnet PITN-GA-2010-264564 
and by
Polish Ministry of Science and Higher Education
   from budget for science for 2010-2013 under grant number N N202 102638.}

\clearpage
\appendix

\section{The experimental cuts\label{app-cuts}}

\subsection{
 $\Phi$ factory KLOE/DA$\Phi$NE (Frascati) -- the reference event
selection 
}

\begin{itemize}

\item $\sqrt{s}$ = 1.02 GeV

\item $E_{\rm min}$ = 0.4 GeV

\item $55 \ensuremath{^\circ} < \theta \pm < 125 \ensuremath{^\circ}$

\item the maximal allowed 2D acollinearity of two charged tracks $\xi_{\rm max}= 9\ensuremath{^\circ}$

\end{itemize}

  To see how the NNLO corrections depend on the event selection we
  obtained results also for

i. wider selection $20 \ensuremath{^\circ} < \theta \pm < 160 \ensuremath{^\circ}$

ii. For each $\theta \pm$  selection the following $\xi_{\rm max}$ was used
 $ (\xi_{\rm max} = 4\ensuremath{^\circ}, 5\ensuremath{^\circ}, 6\ensuremath{^\circ}, 7\ensuremath{^\circ}, 8\ensuremath{^\circ}, 9\ensuremath{^\circ}, 10\ensuremath{^\circ}, 11\ensuremath{^\circ}, 12\ensuremath{^\circ}, 13\ensuremath{^\circ}, 14\ensuremath{^\circ})$
\newline

\subsection{
charm/$\tau$ factory BES-III (Beijing) -- the reference event
selection 
}

\begin{itemize}

\item $\sqrt{s} = $~3.686~GeV, 3.65~GeV and 3.097~GeV

\item $|\cos \theta| < 0.8$, where $\theta$ is the polar angle of the
electron or positron in the lab system, this corresponds to the
barrel region of BES-III detector.
Since in BEPC, $e^+$ and $e^-$
beams are colliding with equal energy but at a 22mrad crossing
angle, the lab system is slightly different from the CoM system.

\item $E_{e^+} > 1.0~{\rm GeV}$ and $E_{e^-} > 1.0~{\rm GeV}$, where $E$ is the energy deposited
in the electromagnetic calorimeter (EMC).\footnote{$e^+$ and $e^-$
deposit virtually all their energy in the EMC, so this is the
energy they carried.}

\end{itemize}
To see how the NNLO corrections depend on the event selection we
  obtained results also for $|\cos \theta| < 0.7,\ 0.75 ,\  0.85 \ {\rm and}\  0.9$.

\vskip 4pt

\subsection{
$B$ factory BaBar/PEP-II (SLAC)  -- the reference event
selection 
}

\begin{itemize}
\item $\sqrt{s}$ = 10.56 GeV

\item $|\vec{p}_{+}|/E_{\rm beam} > 0.75$ and $|\vec{p}_{-}|/E_{\rm beam} > 0.50$

or $|\vec{p}_{-}|/E_{\rm beam} > 0.75$ and $|\vec{p}_{+}|/E_{\rm beam} > 0.50$

\item $|\cos(\theta \pm)| < 0.65$ and $|\cos(\theta +)| < 0.60$ or $|\cos(\theta -) | < 0.60$

\item   the maximal allowed 3D acollinearity of two charged tracks $\xi_{\rm 3dmax} = 30\ensuremath{^\circ}$

\end{itemize}

 To see how the NNLO corrections depend on the event selection we
  obtained results also for

i. $|\cos(\theta \pm)| < 0.70$ and $|\cos(\theta +)| < 0.65$ or $|\cos(\theta -) | < 0.65$

ii. $|\cos(\theta \pm)| < 0.60$ and $|\cos(\theta +)| < 0.55$ or $|\cos(\theta -) | < 0.55$

 For each $|\cos(\theta \pm)|$ selection we have used the following $\xi_{\rm 3dmax}$

$(\xi_{\rm 3dmax} = 20\ensuremath{^\circ}, 22\ensuremath{^\circ}, 24\ensuremath{^\circ}, 26\ensuremath{^\circ}, 28\ensuremath{^\circ} , 30\ensuremath{^\circ}, 32\ensuremath{^\circ}, 34\ensuremath{^\circ},36\ensuremath{^\circ}, 38\ensuremath{^\circ}, 40\ensuremath{^\circ})$
\newline

\subsection{ 
$B$ factory Belle (KEK) -- the reference event
selection 
}

Belle runs at an asymmetric $e^+e^-$ collider,  but
all criteria are expressed in the CoM frame. The selection of Bhabha events by Belle are:

\begin{itemize}

\item $\sqrt{s} = $10.58~GeV 

\item $50.5^\circ<\theta_{\pm}<(180-50.5)^\circ$

\item Two charged tracks have momentum $> 2.645$~GeV

\item The track with maximum deposited energy in EMC greater than 2~GeV,

\item The sum of the deposited energies of all tracks in EMC is greater
than 4~GeV (both charged and neutral particle can deposit energy in EMC and
                              it is not checked if the particle
                              is charged or neutral)

\item Acollinearity angle (2D) $\xi_{\rm 2dmax}< 10^\circ $

\item Transverse momentum of any observed charged particle greater than
0.1~GeV
\end{itemize}
To see how the NNLO corrections depend on the event selection we
  obtained results also for

 i. $45.5^\circ<\theta_{\pm}<(180-45.5)^\circ$

 ii. $55.5^\circ<\theta_{\pm}<(180-55.5)^\circ$

For each $\theta \pm $ selection we have used the following $\xi_{\rm 2dmax}$
$(\xi_{\rm 2dmax} = 5\ensuremath{^\circ}, 6\ensuremath{^\circ}, 7\ensuremath{^\circ}, 8\ensuremath{^\circ}, 9\ensuremath{^\circ} , 10\ensuremath{^\circ}, 11\ensuremath{^\circ}, 12\ensuremath{^\circ},13\ensuremath{^\circ}, 14\ensuremath{^\circ}, 15\ensuremath{^\circ})$

\clearpage

\providecommand{\href}[2]{#2}
\addcontentsline{toc}{section}{References}
\bibliography{2loopsc}

\end{document}